# Evaluating methods for Lasso selective inference in biomedical research by a comparative simulation study

Michael Kammer[1,2], Daniela Dunkler[1], Stefan Michiels[3] and Georg Heinze[1]

1 Medical University of Vienna, Center for Medical Statistics, Informatics and Intelligent Systems, Section for Clinical Biometrics, Vienna, Austria

2 Medical University of Vienna, Department for Internal Medicine III, Division of Nephrology and Dialysis, Vienna, Austria

3 Service de Biostatistique et d'Epidémiologie, Gustave Roussy; INSERM, CESP U1018, University Paris-Saclay, Villejuif, France

## Abstract

Variable selection for regression models plays a key role in the analysis of biomedical data. However, inference after selection is not covered by classical statistical frequentist theory which assumes a fixed set of covariates in the model. We review two interpretations of inference after selection: the full model view, in which the parameters of interest are those of the full model on all predictors, and then focus on the submodel view, in which the parameters of interest are those of the selected model only. In the context of L1-penalized regression we compare proposals for submodel inference (selective inference) via confidence intervals available to applied researchers via software packages using a simulation study inspired by real data commonly seen in biomedical studies. Furthermore, we present an exemplary application of these methods to a publicly available dataset to discuss their practical usability.

Our findings indicate that the frequentist properties of selective confidence intervals are generally acceptable, but desired coverage levels are not guaranteed in all scenarios except for the most conservative methods. The choice of inference method potentially has a large impact on the resulting interval estimates, thereby necessitating that the user is acutely aware of the goal of inference in order to interpret and communicate the results. Currently available software packages are not yet very user friendly or robust which might affect their use in practice. In summary, we find submodel inference after selection useful for experienced statisticians to assess the importance of individual selected predictors in future applications.

## 1. Introduction

Statistical regression models are ubiquitous in the analysis of biomedical data, where they are used to describe associations of an outcome of interest with independent variables, predict the outcome using those independent variables, or to causally explain differences in the outcome by differences in independent variables (1). Modern advances in data collection and measurement technologies facilitate the modelling of more and more details of the underlying multifactorial biological processes. However, to keep the results intelligible and communicable in clinical practice, sparse models are often preferred, which include few covariates selected according to their relationship with the outcome (2). Hence, statistical inference must take the additional uncertainties introduced by such selection procedures into account.

**Inference after selection.** Inference after variable selection, also termed post-selection inference, cannot be performed using classical statistical approaches, and associated problems have been discussed extensively in the literature (see (3-5) for an overview). A key issue of the standard approaches is the assumption of a fixed set of covariates in the model, i.e. that the decisions which parameters to estimate and test are made before observing data. This is no longer valid when the same data is used to perform both of these tasks: first, to select a set of covariates for a model, the so-called active set, and second, to estimate their parameters and provide inference. In this case, the active set itself must be considered as a random component of the model, distinguishing inference after selection from inference with a fixed covariate set. This problem has affected regression modelling since variable selection techniques have been around, i.e. for several decades, but has been mostly ignored for a long time. Increasing availability of data with a larger number variables and increasing usage of variable selection techniques, such as the Lasso, and also increasing awareness of the issues if replicability, have put post-selection inference into the focus of not only methods research, but also applied statistics.

**Inference for the Lasso.** The Lasso is a particular kind of penalized regression which has gained a lot of attention from both theoretical and applied researchers since its introduction in 1996 (6, 7), as it provides automated variable selection and scalability. There are a number of proposals regarding post-selection inference for the Lasso, which address different use-cases and methodological approaches: two-stage approaches (sample splitting and data carving, see e.g. (8-10)), bootstrap based (see e.g. (11-13)), de-sparsified/de-biased Lasso (14-20), approaches controlling the expected number of false positive selections (e.g. Stability selection (21, 22), or knockoff filtering (23)), inference in the presence of many confounders (24, 25), and other conceptual approaches (see e.g. (26-28)). The focus of this work will be on selective inference approaches based on simultaneous inference, and on conditioning on the Lasso selection event (3, 8, 29-31). General overviews are given in Dezeure et al (32) and Hastie et al, chapter 6 (7).

**Objective.** The objective of this paper is to evaluate methods for conducting selective inference in the context of Lasso linear regression in typical biomedical applications. We focus on settings in which the number of observations exceeds the number of variables, where it is realistic to address the task of post-selection inference at the level of single covariates. As of yet there is no independent, comprehensive comparison of approaches available to the applied researcher in practical settings, beyond theoretical considerations and anecdotal data examples (4, 7, 30, 33). This assessment also aims at studying the interpretation of the selective inference framework in realistic usage scenarios.

**Outline.** In the following section we will review key considerations underlying the selective inference framework for regression. We will then introduce the Lasso and several approaches to selective inference. Subsequently, we present the design and discuss the results of our simulation study. A real-data example demonstrates the application of selective inference in practice. We conclude with recommendations for the applied statistician.

# 2. Post-selection inference

## 2.1. Views of post-selection inference

In the classical regression setting the parameters to be estimated and hypotheses to be tested are assumed to be fixed before observing any data. Given a dataset, a model comprising all pre-specified predictors can be estimated, and hypotheses about its regression coefficients targeting the corresponding population parameters can be tested. When conducting variable selection the role of the estimated model is no longer that clear. Berk et al (3) introduced two alternative views of inference after selection, with consequences for the type of research question which can be addressed by statistical inference.

**Full model view.** In the full model view, the population parameters of interest are those of the full set of candidate predictors, possibly including interactions or non-linear terms. The “full” model comprising all candidate predictors is the object of interest for future research and is assumed to be a (possibly causally interpreted) description of the data generating mechanism.

Variable selection therefore merely amounts to forcing some of the coefficients in the estimated model to zero, but a corresponding population parameter still exists and serves as the target of post-selection inference. Approaches based on selection probabilities or e.g. the de-sparsified Lasso adopt this interpretation.

**Submodel view.** In the submodel view, interest lies in the parameters of the selected variables only. The full model does not have a special meaning, as a models' primary purpose is to provide a succinct description of the association of the outcome and the independent variables, not necessarily capturing the data generating mechanism. The hypotheses to be assessed, and also the population quantities, therefore depend on the selected submodel, giving rise to the term *selective inference* (4, 8, 30, 34). This interpretation is adopted by the methods evaluated in this paper.

**Practical implications.** In exploratory studies, researchers may be interested if the effect of a covariate is truly expected to be small in the population, even if set to zero in the estimated model. This naturally aligns with the full model view of inference, which provides inference for all candidate predictors entering the selection process. Valid inference then necessitates strong assumptions regarding the correctness of the full model, and the true sparsity of its parameters. For example, a common condition is that all non-zero regression coefficients are larger than a certain noise threshold. In practice it can seldom be ruled out that weak predictors are present in the data generating mechanism, and several results indicate that the biases incurred by variable selection through misspecification of the selected model, e.g. omitted variable bias, make general inference for the true underlying parameters very challenging (35, 36). On the other hand, researchers may not be interested in all parameters, but are willing to focus future efforts on a smaller subset of them. In this case, the submodel approach circumvents the strong assumptions of full model inference by restricting inference to the selected variables. In contrast to classical theory, the submodel view need not assume that the submodel is correct, i.e. estimates the conditional expectation of the outcome, but the submodels' linear predictor is merely regarded as an approximation of said expectation. For details we refer to Section 2.2 in Berk et al (3). A graphical illustration of the two views is provided in Figure 1.

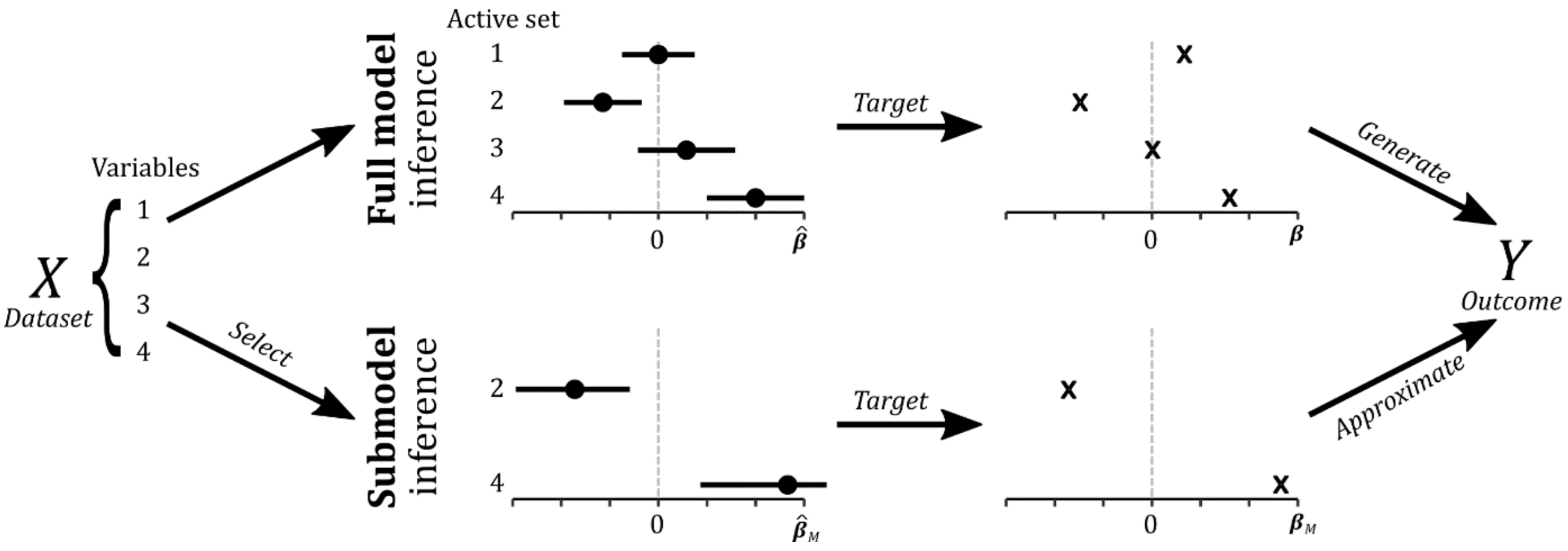

Figure 1: Graphical illustration of the two views of post-selection inference. Under the full model view, each candidate predictor (even if the estimated coefficient is zero) has a target parameter for inference. In the submodel view, only the selected predictors have defined population parameters, and these may differ (in case of correlation between predictors) from the full model targets. Interval estimates, indicated by black lines, are also different (likely wider) between full model and submodel inference due to the need to account for variable selection.

## 2.2. Selective inference

Selective inference is a general paradigm to address issues arising when the hypotheses of interest are not specified before data collection, but defined during the process of data analysis. A prototypical situation concerns the use of variable selection procedures to select important predictors. Valid inference can then be derived from the key insight that when the mechanism of selection is known, the resulting selection bias can often be mitigated (5).

**Notation.** We adopt the notation of Berk et al (3) and denote a model by the variable it comprises (i.e. its active set). Thus, if the full set of candidate predictors consists of $p$ variables, we write $M_F := \{1, 2, \dots, p\}$ for the full model using all predictors, and $M \subseteq M_F$ for any (fixed) submodel. We use the notation $\widehat{M}(\boldsymbol{y})$ (generally abbreviated as $\widehat{M}$) for the model chosen by a specific variable selection procedure, depending on the outcome $\boldsymbol{y}$. The vector of regression coefficients corresponding to a specific choice of predictors $M$ is denoted as $\boldsymbol{\beta}_M := \left(\beta_{j,M}\right)_{j \in M}$ (if $M = M_F$ we will omit the index). Similarly, we write $\boldsymbol{X}$ for the full set of observations of dimension $n \times p$ and $\boldsymbol{X}_M$ for the dataset comprising only the variables in $M$.

**General selective inference.** In classical statistical inference, a confidence interval (CI) $CI_j$ for an estimated regression coefficient $\widehat{\beta_j}$ is defined as a contiguous set of numbers such that the probability $\mathbb{P}\left[\beta_j \in CI_j\right]$ to cover the true parameter $\beta_j$ is $1 - \alpha$, where $\alpha$ is the pre-specified significance level. In analogy, selective CIs can be defined by model-dependent coverage probabilities. These apply when the researcher conducts variable selection to obtain a model $M$, and subsequently is interested in inference for the variables in $M$.

**Selective inference for the submodel view.** A central quantity in the definition to all the methods in this manuscript are the submodel population regression coefficients $\boldsymbol{\beta}_M$ specific to a given model $M$. In the case of linear regression these are defined through the requirement of unbiasedness and are estimable linear functions of the full model population parameters $\boldsymbol{\beta}$, such that $\boldsymbol{\beta}_M = \mathbb{E}(\widehat{\boldsymbol{\beta}}_M) = (\boldsymbol{X}_M^T \boldsymbol{X}_M)^{-1} \boldsymbol{X}_M^T \boldsymbol{X} \boldsymbol{\beta}$ (see Berk et al. (3), formula 3.2)**.** Note, however, that $\boldsymbol{\beta}_M$ is not assumed to be an estimate for the corresponding full model parameters in $\boldsymbol{\beta}$**.**

**Conditional selective inference.** In practice, one could also be interested in inference for an a-priori fixed variable of interest $j$. This requires conditioning the selective coverage probabilities on the event that the variable is included in the selected model, otherwise inference is not of interest. Such situations occur, for example, when one adds a new candidate predictor to a pool of known variables, performs variable selection to sparsify the model, and then conducts inference for the new candidate, given that it survived the selection procedure. However, coverage guarantees of this kind depend on the selection probability of the variable, thus leading to reliable inference only for strong predictors (3).

# 3. Methods

## 3.1. The Lasso regression family

Many biomedical questions may be analysed by generalized linear models (GLM). These assume that the outcome variable of interest $Y$ is generated by a distribution with density $f$ from an exponential family (including e.g. the normal or binomial distributions). The conditional expectation $\mathbb{E}(Y|\boldsymbol{X})$ is modelled depending on the observed, fixed data $\boldsymbol{X}$ by the general functional form

$$g\big(\mathbb{E}(Y|\boldsymbol{X})\big) = \beta_0 + \boldsymbol{X}\boldsymbol{\beta}.$$

Here $g$ is a link function, $\beta_0$ is an (optional) intercept term and $\boldsymbol{\beta}$ is a vector of regression coefficients. The transformation $g$ between the linear predictor $\boldsymbol{X}\boldsymbol{\beta}$ and $Y$ is chosen according to the type of the outcome variable. In this work we are mainly concerned with the continuous response Gaussian regression model, where $g$ is the identity. The general concepts outlined for selective inference apply to the whole family of models, however. The regression coefficients can generally be obtained by maximising the likelihood function $L(\boldsymbol{\beta}) = f(\boldsymbol{y}|\boldsymbol{X}, \boldsymbol{\beta})$ given the observed data $\boldsymbol{X}$ and $\boldsymbol{y}$.

The Lasso regression model can be obtained by changing the maximised objective function to a penalized likelihood

$$\mathrm{L}_P(\boldsymbol{\beta}) = L(\boldsymbol{\beta}) - \lambda \sum_i |\beta_i|,$$

where $\lambda$ constitutes a tuning parameter controlling the impact of the penalization term, given by the $l_1$ norm of the regression coefficients (6). Prediction accuracy is balanced against model complexity by restricting the set of possible solutions to the optimization problem, thereby addressing issues with standard maximum likelihood estimation, such as collinearity. This special form of penalization may induce variable selection by forcing some of the entries of the estimated $\widehat{\boldsymbol{\beta}}$ to exactly zero. The Lasso objective is convex, such that efficient algorithms exist to compute estimates for a whole path of $\lambda$ values (37). Thanks to these two features of sparsity and computational accessibility the Lasso was widely adopted in many fields of modern science. However, the Lasso suffers from several drawbacks, such as a high false positive selection rate and a bias towards zero for large coefficients (7, 38, 39).
The adaptive Lasso (40) addresses these issues by introducing weights in the penalty term $\sum_i w_i|\beta_i|$. These pre-specified weights allow differential penalization of the individual predictors, thereby allowing weak predictors to undergo stronger restrictions than strong predictors. Weights can be obtained from e.g. the unpenalized GLM coefficients as $w_i = 1/\left|\beta_{i,GLM}\right|^\gamma$, using a second tuning parameter $\gamma$ to control the impact of the weights on the penalization. The adaptive Lasso enjoys a so-called oracle property which the ordinary Lasso lacks, namely it offers consistent variable selection under conditions that can be considered realistic with large sample sizes. Thus, it is able to identify the predictors of the "true" data generating model with high probability, if they are part of the set of candidate predictors (40, 41). The adaptive Lasso can be easily implemented in any software that is able to fit the ordinary Lasso by re-scaling the input data: weighting the contributions to the penalty term of individual coefficients $\left|\beta_j\right|$ by $w_j > 0$ is equivalent to scaling the corresponding column in $\boldsymbol{X}$ by $1/w_j$.

## 3.2. Selective inference for the Lasso

### 3.2.1. Sample splitting (Split)

Sample splitting is an intuitive approach to selective inference agnostic to the model selection procedure, introduced already in 1975 (42). It consists of partitioning the dataset in two (not necessarily equally sized) parts. First, a set of active variables $M$ is derived from one part of the dataset. Inference is then based only on the other part of the dataset, in which the set of active variables can be considered fixed, conditional on the data used in the first step. Thus, classical statistical theory can be applied to obtain selective inference for $M$. This approach controls the submodel coverage at the nominal significance level $\alpha$ such that $\mathbb{P}\left[\beta_{j,M} \in CI_{j,M} | \widehat{M} = M\right] \geq 1 - \alpha$ for $j \in M$. The uncertainty introduced by selection is reflected through the part of the data that is not available for inference computations. Sample splitting is easily implemented in any statistical software. The two parts of the dataset can be of unequal sizes, related to a trade-off between selection and inference accuracy. Simulations suggest that a simple 50%-50% split offers a good compromise (8, 28).

### 3.2.2. Exact post-selection inference for the Lasso (SI)

This procedure proposed by Lee et al (30) constructs selective CIs that guarantee coverage at the nominal significance level $\alpha$, conditional on the specific model $M$ that was selected by the Lasso:

$$P\left[\beta_{j,M} \in CI_{j,M} |\ \widehat{M} = M\right] \geq 1 - \alpha, j \in M.$$

The authors show that the event of selecting a model $M$ by the Lasso corresponds to a polyhedral region in the space $\mathbb{R}^p$ of regression coefficients, and are thereby able to analytically derive the sampling distribution conditional on $M$ required to compute CIs. Point estimates $\widehat{\boldsymbol{\beta}}_M$ can be obtained by fitting a GLM on the active set $M$. In the case of continuous outcome regression, the procedure assumes an independent estimate of the outcome variance $\sigma^2$ to provide valid inference. Because of the correspondence between weighting and scaling the input data, the same algorithm can also be used to obtain selective inference for the adaptive Lasso. Notably, this method was derived for the special case of the Lasso with a fixed tuning parameter $\lambda$. In practice, this is not a realistic usage scenario as the penalization strength is most likely tuned via cross-validation or information criteria. Computer intensive extensions to allow tuning of $\lambda$ have been developed, but are not yet available as a software package (28, 43). The worst-case coverage of the proposed intervals is poor when the Lasso estimator is tuned (4) but there are also comments that the effect of tuning on the intervals is likely small (7). Further extensions are aimed at increasing the power of the approach by conditioning only on the selected variables, and dropping the conditioning on the signs of their estimated coefficients (31).

#### 3.2.3. Universally valid post-selection inference (PoSI)

In contrast to the conditional coverage of Lee et al, the approach by Berk et al (3) guarantees valid CIs irrespective of model selection strategy. For a given significance level $\alpha$, the authors propose to control the family wise error rate

$$\forall \widehat{M} \subseteq M_F\colon\ \mathbb{P}\left[\forall j \in \widehat{M}\colon \beta_{j,\widehat{M}} \notin CI_{j,\widehat{M}}\right] \leq \alpha.$$

Given $p$ candidate predictors, their procedure re-casts selective inference as a multiple testing problem. For any specific selected submodel $M$ and its estimated coefficients $\widehat{\boldsymbol{\beta}}_M$, symmetric confidence intervals are formed as $CI_{j,M} = \left[\hat{\beta}_{j,M} \pm K\ \hat{\sigma}(\hat{\beta}_{j,M})\right]$, where $\hat{\sigma}(\hat{\beta}_{j,M})$ denotes an independently estimated standard error of the $j$-th entry of the coefficient vector $\widehat{\boldsymbol{\beta}}_M$. The "PoSI" multiplier $K$ is computed to account for the selection of $M$ from the space of all submodels of $M_F$. It depends on the correlation structure of the dataset**,** desired coverage and an independent estimate of the outcome variance $\sigma^2$. As there is no general closed-form expression for $K$, it is approximated by Monte-Carlo simulation. In practice, the point estimates $\widehat{\boldsymbol{\beta}}_M$ and their associated standard errors can be obtained by fitting a linear regression model on the variables in $M$. This method, which so far has only been developed for continuous outcome regression models, yields simultaneous error control for every predictor and every possible submodel. It is therefore expected to be a very conservative procedure, but recent extensions may improve efficiency of the method (4). Furthermore, the necessary computational resources can become prohibitive with a large number of candidate predictors. It is possible to restrict the adjustment of the CIs to a subset of all submodels, e.g those containing a fixed maximum number of variables only. An advantage is that the PoSI method can be used with any kind of model selection procedure, e.g. the Lasso with tuned penalization strength.

## 4. Simulation design

A brief summary in form of a simulation profile as well as details regarding the simulation design, which is reported following Morris et al (44), can be found in the Supplementary Material.

### 4.1. Aim

The aim of this simulation study was to evaluate recent proposals for selective inference in the context of Lasso regression regarding their frequentist properties in practical usage scenarios.

## 4.2. Data generation

We restricted this simulation study to low-dimensional settings commonly seen in biomedical data analysis, in which the number of observations is larger than the number of variables, but their ratio may be below the often cited threshold of 10 to 15 observations per variable (45). Our guiding design principle for data generation was to study how correlation structures and distribution of effects affect the frequentist properties of selective CIs.

Our general simulation workflow is depicted in Figure 2. First, we sampled the covariates $\boldsymbol{Z}$ from a multivariate normal distribution with pre-specified correlation structure. The resulting values were then transformed (denoted here by $\boldsymbol{T}$) to obtain different predictor distributions. Given the final data matrix $\boldsymbol{X} = \boldsymbol{T}(\boldsymbol{Z})$ of pre-specified sample size, we computed the true outcome $\boldsymbol{y} = \boldsymbol{X\beta} + \boldsymbol{\epsilon}$ using a pre-specified coefficient vector $\boldsymbol{\beta}$ for the linear predictor. The signal-to-noise ratio (target coefficient of determination $R^2$) was controlled via the variance of the error term $\boldsymbol{\epsilon}$ drawn from a normal distribution with mean zero. Validation data was obtained by using the same realisation of the data matrix $\boldsymbol{X}$ and drawing another error vector $\boldsymbol{\epsilon}'$ to obtain a new outcome vector $\boldsymbol{y}'$.

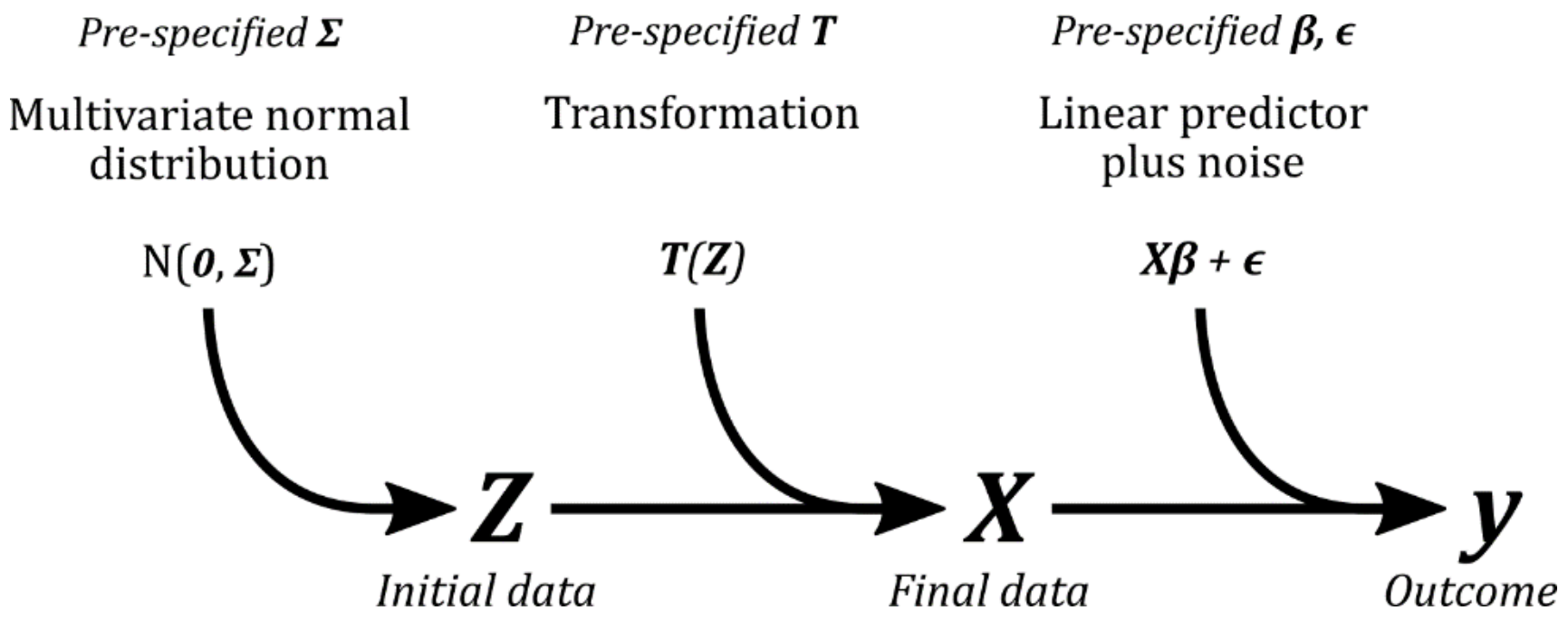

Figure 2: Overview of data generation procedure.

Using this generic simulation procedure, we created two different setups: a simple 'toy setup' with multivariate normal data (the transformation $\boldsymbol{T}$ is simply the identity function) using various correlation structures, and a 'realistic setup' with more realistic distributions and dependencies. In the latter case, the transformation $\boldsymbol{T}$ uses affine and exponential functions, as well as thresholding to yield continuous and binary variables based on a simulation setup by Binder et al (46). The setup includes skewed distributions that can pose problems to statistical approaches, but are hardly considered in simulation studies.

## 4.3. Estimands

The estimands of interest were the expected frequentist properties of selective CIs and conditional selective CIs for regression coefficients obtained after variable selection, excluding the intercept terms. The confidence level was fixed at 90%, the default level in the main software package used. Our evaluation was focused on the following three estimands (Table 2):

- Selective coverage: how often is the submodel target parameter covered by the CI?
- Selective power: for a variable with submodel target parameter unequal to zero, how often does its CI exclude zero?
- Selective type 1 error: for a variable with submodel target parameter equal to zero, how often does the CI exclude zero?

| | | **Toy setup** | **Realistic setup** |
|---|---|---|---|
| **Fixed design parameters** | Motivation | Simplicity, insight | Realistic data |
| | Regression type | Gaussian | Gaussian |
| | Number of variables | 4 | 17 |
| | Type of variables | Continuous | Continuous, binary |
| | Distribution of variables | Gaussian | Mixed (specified in Supplementary Table S 2) |
| **Varying design parameters** | Correlation structures $\boldsymbol{\Sigma}$ | 7 blocked correlation matrices with no or strong correlation | Fixed, mimicking real study (Supplementary Figure S 2) |
| | Coefficient structures $\boldsymbol{\beta}$ | 10 (Supplementary Table S 1) | 13 (Supplementary Table S 3) |
| | True target $R^2$ (noise $\boldsymbol{\epsilon}$) | 0.2, 0.5, 0.8 | 0.2, 0.5, 0.8 |
| | Observations per variable | 5, 10, 50 | 5, 10, 50 |
| **Simulation parameters** | Number of scenarios | 630 | 117 |
| | Iterations per scenario | 900 | 900 |

Table 1: See the description of the simulation setups in the Supplementary Material for more details on the correlation and coefficient structures. Varying design parameters are the parameters which were varied in the full factorial design in both setups. The notations $\boldsymbol{\Sigma}$ and $\boldsymbol{\beta}$ correspond to Figure 2.

The general frequentist properties were obtained by marginalizing (averaging) over all selected models and variables in those models. In the case of conditional inference the marginalization was only over all selected models comprising a specific variable. The submodel targets $\boldsymbol{\beta}_{\widehat{M}}$ depend on the selected model $\widehat{M}$ and may be different in each iteration of the simulation and for each variable selection method. They were computed using the true covariance matrix and the pre-specified population values of the full vector of regression coefficients $\boldsymbol{\beta}$. In particular, $\boldsymbol{\beta}_{\widehat{M}} = \boldsymbol{\Sigma}_{\widehat{M}}\boldsymbol{\Sigma}_{\widehat{M},M_F}\boldsymbol{\beta}$, where $\boldsymbol{\Sigma}_{\widehat{M}}$ is the $q \times q$ submatrix of the true covariance matrix $\boldsymbol{\Sigma}$ which contains only the variables in $\widehat{M}$, and $\boldsymbol{\Sigma}_{\widehat{M},M_F}$ is the $q \times p$ submatrix which contains all the covariances of the active variables in $\widehat{M}$ and the whole set of candidate predictors $M_F$.

**Secondary estimands.** To further aid in the interpretation of the inference properties, we report simulation result for several other estimands:

- Probability of true model selection, estimated by the relative frequency with which the true model (i.e. the true data generating model underlying the simulated data) was selected
- Variable selection probability, estimated by the relative frequency with which a specific variable was selected
- Median width of selective CIs
- Predictive accuracy, in terms of $R^2$ on validation data

**Inference stability.** Since the SI method has been shown to lead to CIs of extreme or even infinite width in some cases (30, 33), we also report the relative frequency with which this occurs.

| | Selective estimand | Definition | Approximation by simulation |
|---|---|---|---|
| **General selective inference** | Coverage | $\mathbb{P}\left[\beta_{.,\widehat{M}} \in CI_{.,\widehat{M}}\right]$ | $\frac{\sum_{j\in M_F}\sum_{M\subseteq M_F}\sum_{s\in S}\mathbb{I}\left[\widehat{M}_s = M \wedge \beta_{j,M} \in CI_{j,M}\right]}{\sum_{j\in M_F}\sum_{s\in S}\mathbb{I}\left[j \in \widehat{M}_s\right]}$ |
| | Power | $\mathbb{P}\left[\beta_{.,\widehat{M}} \in CI_{.,\widehat{M}} \middle\vert \beta_{.,\widehat{M}} \neq 0\right]$ | $\frac{\sum_{j\in M_F}\sum_{M\subseteq M_F}\sum_{s\in S}\mathbb{I}\left[\widehat{M}_s = M \wedge\ 0 \notin CI_{j,M}\right]}{\sum_{j\in M_F}\sum_{s\in S}\mathbb{I}\left[j \in \widehat{M}_s \wedge \beta_{j,\widehat{M}_s} \neq 0\right]}$ |
| | Type 1 error | $\mathbb{P}\left[\beta_{.,\widehat{M}} \in CI_{.,\widehat{M}} \middle\vert \beta_{.,\widehat{M}} = 0\right]$ | $\frac{\sum_{j\in M_F}\sum_{M\subseteq M_F}\sum_{s\in S}\mathbb{I}\left[\widehat{M}_s = M \wedge\ 0 \notin CI_{j,M}\right]}{\sum_{j\in M_F}\sum_{s\in S}\mathbb{I}\left[j \in \widehat{M}_s \wedge \beta_{j,\widehat{M}_s} = 0\right]}$ |
| **Conditional selective inference** | Coverage | $\mathbb{P}\left[\beta_{j,\widehat{M}} \in CI_{j,\widehat{M}} \middle\vert j \in \widehat{M}\right]$ | $\frac{\sum_{M\subseteq M_F}\sum_{s\in S}\mathbb{I}\left[\widehat{M}_s = M \wedge \beta_{j,M} \in CI_{j,M}\right]}{\sum_{s\in S}\mathbb{I}\left[j \in \widehat{M}_s\right]}$ |
| | Power | $\mathbb{P}\left[\beta_{j,\widehat{M}} \in CI_{j,\widehat{M}} \middle\vert j \in \widehat{M}, \beta_{j,\widehat{M}} \neq 0\right]$ | $\frac{\sum_{M\subseteq M_F}\sum_{s\in S}\mathbb{I}\left[\widehat{M}_s = M \wedge\ 0 \notin CI_{j,M}\right]}{\sum_{s\in S}\mathbb{I}\left[j \in \widehat{M}_s \wedge \beta_{j,\widehat{M}_s} \neq 0\right]}$ |
| | Type 1 error | $\mathbb{P}\left[\beta_{j,\widehat{M}} \in CI_{j,\widehat{M}} \middle\vert j \in \widehat{M}, \beta_{j,\widehat{M}} = 0\right]$ | $\frac{\sum_{M\subseteq M_F}\sum_{s\in S}\mathbb{I}\left[\widehat{M}_s = M \wedge\ 0 \notin CI_{j,M}\right]}{\sum_{s\in S}\mathbb{I}\left[j \in \widehat{M}_s \wedge \beta_{j,\widehat{M}_s} = 0\right]}$ |

Table 2: Definition of selective estimands investigated. We denote the set of all iterations of a simulation scenario by $S = \{1, \dots, n_{sim}\}$, where $n_{sim}$ is the total number of iterations. The full model using all predictors is written as $M_F = \{1, \dots, p\}$, the selected model in a specific iteration $s$ is written as $\widehat{M}_s$. By the use of $\mathbb{I}[.]$ we denote the indicator function for the event specified between square brackets. Note that for methods without variable selection, general and conditional estimands coincide and reduce to the usual definitions of frequentist properties. More details on the derivation of the approximation in the simulation are given in the Supplementary Material.

## 4.4. Methods

An overview of all methods included in the comparison is provided in Table 3.

**Variable selection procedures.** Variable selection in our study was based on the Lasso (Lasso) and the adaptive Lasso (ALasso). We used the reciprocals of the absolute values of the coefficient estimates from the model using all predictors (full model) as penalization weights in the adaptive Lasso.
The penalization parameters were determined in one of two ways, which differ in whether or not the observed outcomes were used in the estimation process. By doing so, we could compare the impact of tuning the penalization strength instead of using a fixed penalization parameter, an important issue for the SI method.

- Cross-validation (CV): use 10-fold cross validation to obtain the penalization parameter with minimal estimated prediction error. The observed outcomes $\boldsymbol{y}$ directly affect the determined $\lambda$, which is therefore a random variable itself.
- Estimation following Negahban et al (Neg, (47)): this procedure relies on the data $\boldsymbol{X}$ and an estimate of the marginal outcome variance $\sigma^2$, and is therefore not specific to the observed outcomes $\boldsymbol{y}$. The penalization parameter is estimated as $\lambda = 2\mathbb{E}\left(\left\|\boldsymbol{X}^T\epsilon\right\|_\infty\right)$, where $\epsilon \sim N(0, \sigma^2)$. Thus, $\lambda$ can be considered as a fixed, non-random parameter. This method was used only in combination with SI.

**Inference procedures.** We evaluated the following selective inference procedures after variable selection with the Lasso or adaptive Lasso:

- Sample splitting (Split): random split of the dataset into two halves for selection and inference
- Exact post-selection inference (SI, (30))
- Universally valid post-selection inference (PoSI, (3))

**Reference procedures.** For comparison, we included two methods without data-driven variable selection: first, the full model (Full), which was always estimable in our low-dimensional setting. Second, the oracle model (Oracle), which included only the true predictors used for generating the data. The latter reflects the ideal but unrealistic situation that is assumed by classical statistical inference of perfect knowledge of the variables involved in the data generating mechanism. However, coefficient estimates for both reference models are still subject to finite sample variability.

| Method | Variable selection | Tuning | Inference |
|---|---|---|---|
| Full | None | None | Wald CI |
| Oracle | None | None | Wald CI |
| Lasso-CV-Split | Lasso | 10-fold CV | Split-sample |
| Lasso-CV-PoSI | Lasso | 10-fold CV | Universally valid post-selection inference (3) |
| Lasso-CV-SI | Lasso | 10-fold CV | Exact post-selection inference (30) |
| Lasso-Neg-SI | Lasso | Fixed penalization parameter (47) | Exact post-selection inference (30) |
| ALasso-CV-Split | Adaptive Lasso | 10-fold CV | Split-sample |
| ALasso-CV-PoSI | Adaptive Lasso | 10-fold CV | Universally valid post-selection inference (3) |
| ALasso-CV-SI | Adaptive Lasso | 10-fold CV | Exact post-selection inference (30) |
| ALasso-Neg-SI | Adaptive Lasso | Fixed penalization parameter (47) | Exact post-selection inference (30) |

Table 3: Overview of methods investigated in this study.

## 4.5. Performance measures

For the selective coverage and type 1 error rate, we evaluated the methods by their discrepancy to nominal significance levels for the estimated CIs. For selective power, the method with the highest value was considered to be the best performing method. For variable and model selection frequencies, methods closer to the true oracle model were considered better. Predictive accuracy, or relative prediction performance of the models was measured by the absolute bias of the achieved $R^2$ on validation data and the true $R^2$.

## 4.6. Software and implementation details

All analyses were implemented in the R statistical software, version 3.5.1 (48). We used the packages *glmnet* (49) (version 2.0-18), for implementing Lasso and ALasso, *selectiveInference* (50) (version 1.2.4) for the SI and *PoSI* (51) (version 1.0) for the PoSI method. Data simulation and visualisation was facilitated by the *simdata* (52) and *looplot* (53) packages.
The selectiveInference algorithm involved a grid search for the bounds of the selective CIs for which the search space was extended from the default values to [-1000, 1000] with 1000 interval steps. Some minor bugs were fixed in the package to ensure correct computation of the CIs (see corresponding section in Supplementary Material).
The algorithm for the computation of the PoSI constant as implemented in the PoSI package involved numeric simulations. Due to the computational burden, we reduced the number of these internal simulations to 500 for the realistic setup only.
We left all other options for the selectiveInference and PoSI packages at their default values.
In-line with the literature, we derived estimates of the outcome variance $\hat{\sigma}^2$ from the residuals of the full model. For the Neg method we approximated the expectation required for the estimation of $\lambda$ by 1000 internal simulations, using $\epsilon \sim N(0, \hat{\sigma}^2)$. All computations were done on standardized covariate data such that all columns in the matrix $\boldsymbol{X}$ had a mean of zero and unit variance.

# 5. Results

In this section we will mainly focus on the Toy setup to explain the results from the simulation study and detail the differences to the realistic setup whenever necessary. Generally, results from the Toy setup were more variable due to the wide variety of setups studied, as compared to the Realistic setup.

## 5.1. Model and variable selection

Figure 3 and Supplementary Figure S 3 give an overview of the results for the toy and realistic setups, respectively. The frequencies of correct model selection were much lower in the realistic setup than in the toy setup, due to the larger number of variables in the realistic setup, leading to a higher false positive rate.
Regarding the frequency of true model selection, the ALasso-CV approaches clearly outperformed the Lasso-CV approaches. The ALasso-CV method selected the true model more often (see upper row in Figure 3) and included false positives (noise variables) less often in scenarios with higher signal-to-noise ratios (see bottom row in Figure 3). In contrast, the false positive rate increased for Lasso-CV with higher target $R^2$. Lasso-Neg and ALasso-Neg generally led to very sparse models, with good properties regarding false positives, but at the cost of a higher probability of missing important predictors. Lasso-CV-Split and ALasso-CV-Split resulted in slightly lower accuracy of model selection due to less data being available for the selection. This was especially noticeable in the case of the ALasso-CV-Split, most likely because of inaccurately estimated penalization weights in the reduced subsample. Exemplary results for the toy setup regarding individual variable selection frequencies are shown in Supplementary Figure S 4.

## 5.2. Selective inference

### Stability

The stability of the Lasso-CV-SI method was a concern in our simulations. The main issue were CIs which did not include the point estimate of the regression coefficient. In such cases, the width of the CIs was extremely sensitive regarding the inclusion of certain (weak) predictors, leading to highly variable results. Almost 26% of the iterations of the realistic setup resulted in unstable CIs (2% had CIs with infinite width, 26% had CIs excluding the point estimate), while this happened only for 5% of the runs in the smaller toy setup. This issue was worst when all predictors of the full model were strongly positively correlated. Its frequency increased with the signal-to-noise ratio and sample size due to the Lasso’s tendency to include weak predictors. Note that whenever even a single weak predictor was included, inference for the whole model was problematic due to extremely wide CIs for all included variables. The problem was much less severe for ALasso-CV-SI (less than 1% of runs with infinite CIs, 6% with CIs excluding the point estimate for the realistic setup) due to its better variable selection properties. For this method, unstable CIs became less frequent with increasing signal strength and sample size. In the case of Lasso-Neg-SI and ALasso-Neg-SI stability was also found to be less problematic (less than 4% and 3% of iterations with unstable CIs, respectively).

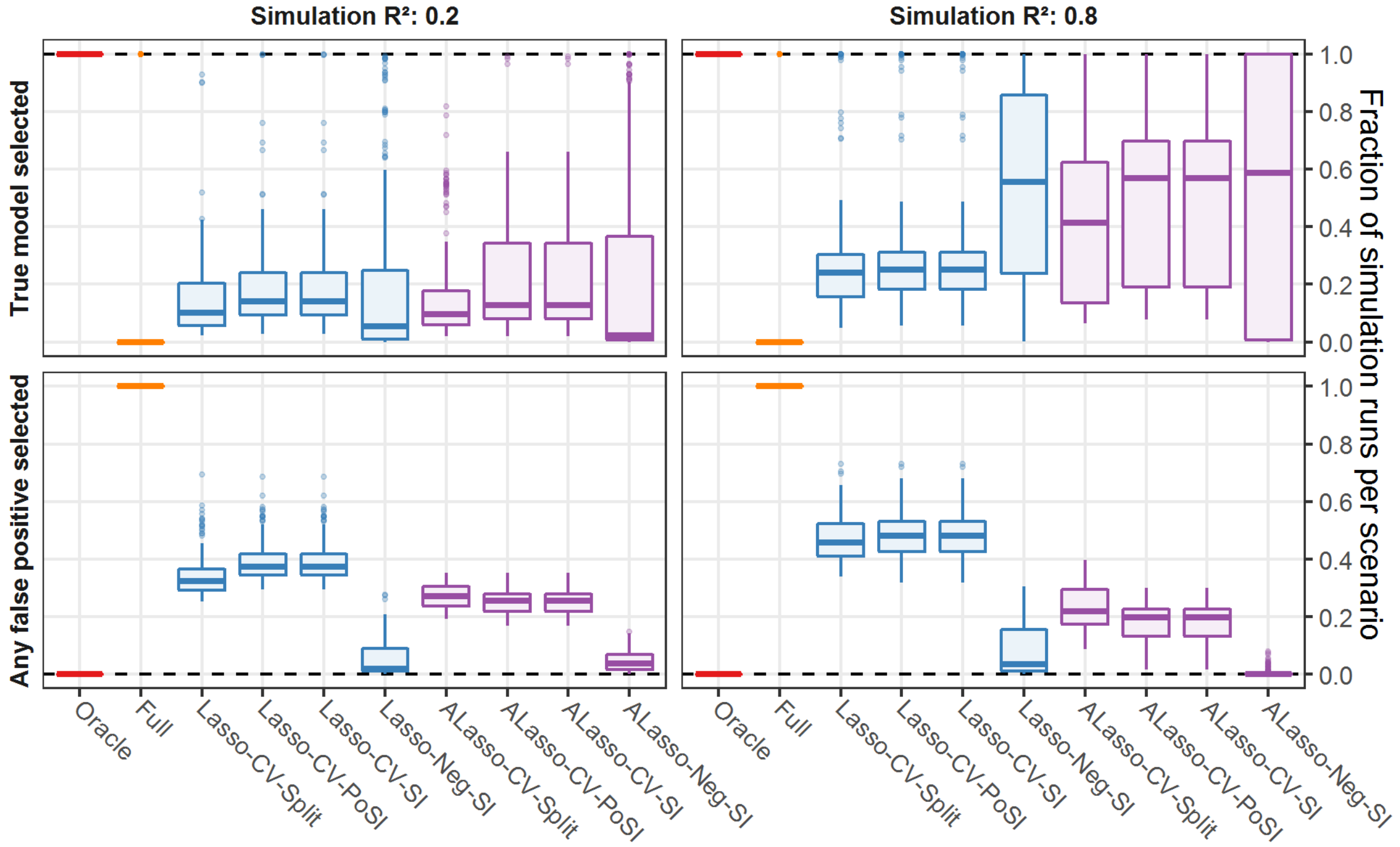

Figure 3: Summary of results from the toy setup regarding model selection (only the two extreme $R^2$ values are depicted; for $R^2 = 0.5$ the results are in between). For each scenario, we computed the fraction of simulation runs in which the true data generating model (i.e. only true predictors) was selected (top row) and in which any false positive selection (i.e. a true non-predictor was selected) was made (bottom row panel). The boxplots summarise these results over all scenarios. The target, optimal values are depicted as dashed lines. Colors are based on the type of regression model used.
Results indicate that the probability to selected the true model increases for the ALasso approaches, but remains constant for the Lasso approaches. Similarly, the probability to select false positives decreases for ALasso, but increases slightly for the Lasso.

### Primary estimands

Summary results for selective coverage in both simulation setups are shown in Figure 4. Selective power and type 1 error are displayed in Figure 5 and Supplementary Figure S 5 for the toy and realistic setups, respectively. These results were based on all runs per scenario. Selective coverage and type 1 error remained stable across the different signal-to-noise ratios and were similar between both simulation setups, while selective power increased with higher signal-to-noise ratio. Using Lasso-CV-SI led to slightly lower-than-nominal coverage properties (in median 0.025 below target for the toy setup, less than 0.01 for the realistic setup). As expected, Lasso-CV-PoSI and ALasso-CV-PoSI were conservative, especially in the realistic setup with a larger number of variables, and yielded selective type 1 error rates largely below 0.05. In contrast to the Lasso, ALasso-CV-SI consistently led to noticeable lower-than-nominal selective coverage (in median 0.06 below target in the toy setup). While the selective power was higher than for the Lasso, the number of false positive selective significances increased drastically. Lasso-CV-Split and ALasso-CV-Split led to similar results, as sample splitting is agnostic of the model selection procedure.
In the realistic setup a relevant proportion of runs led to unstable CIs by the SI method. Removing the unstable runs from the analyses shifted results towards overcoverage, especially in the case of Lasso-CV-SI for which 26% of all iterations were deemed unstable: median selective coverage over all scenarios increased from 0.89 to 0.93, median selective type 1 error decreased from 0.12 to 0.06.

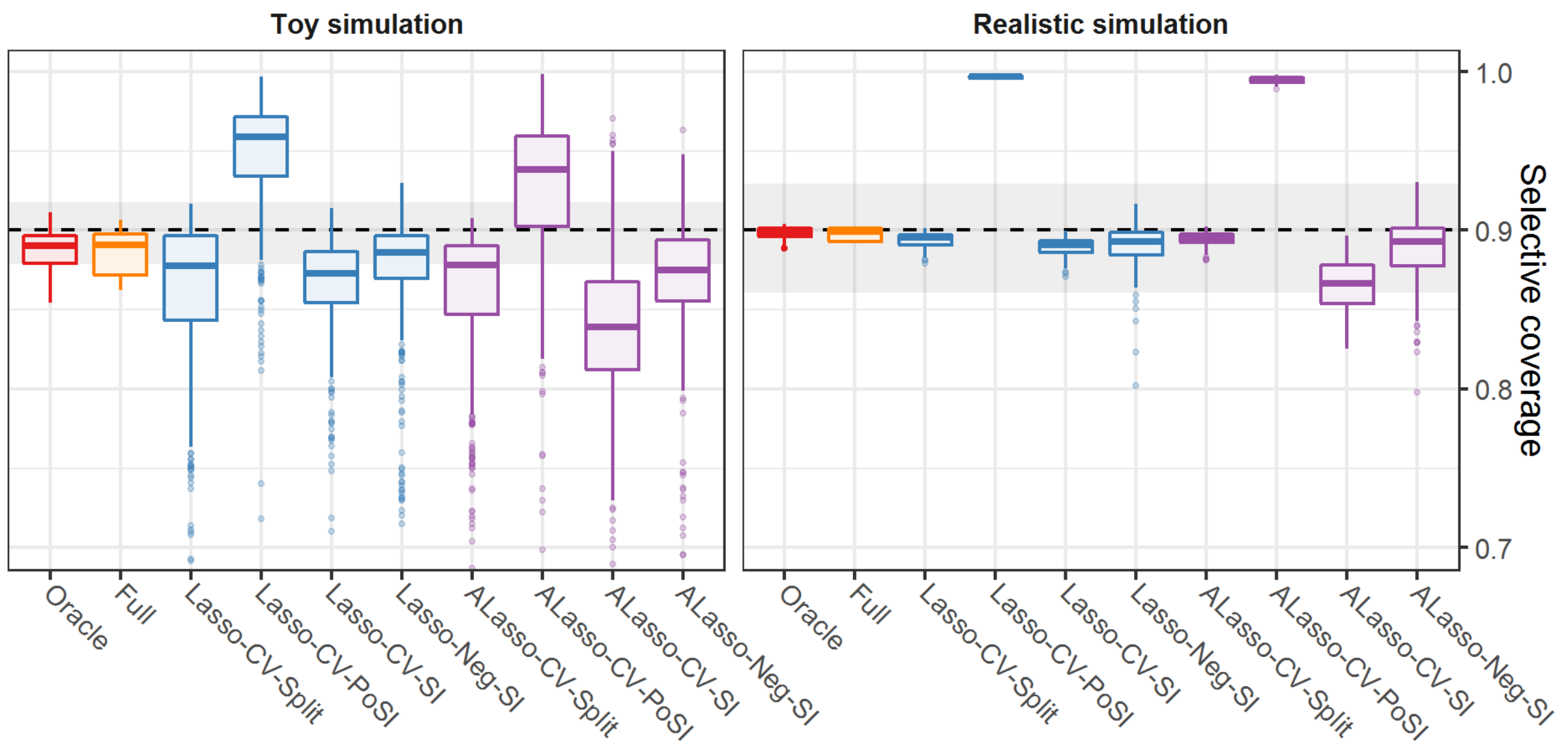

Figure 4: Summary of results from both simulation setups (toy setup in left panel, realistic setup in right panel) regarding selective coverage of the selective 90% CIs for the submodel inference target (see Table 2). Results for all scenarios are summarised by boxplots. The nominal confidence level of 0.9 used in the construction of the CIs is depicted as dashed lines. Colors are based on the type of regression model used. To give an indication of variability expected in this simulation study, the grey areas indicate binomial 95% confidence intervals based on the number of iterations in each scenario (900) and the nominal confidence level.

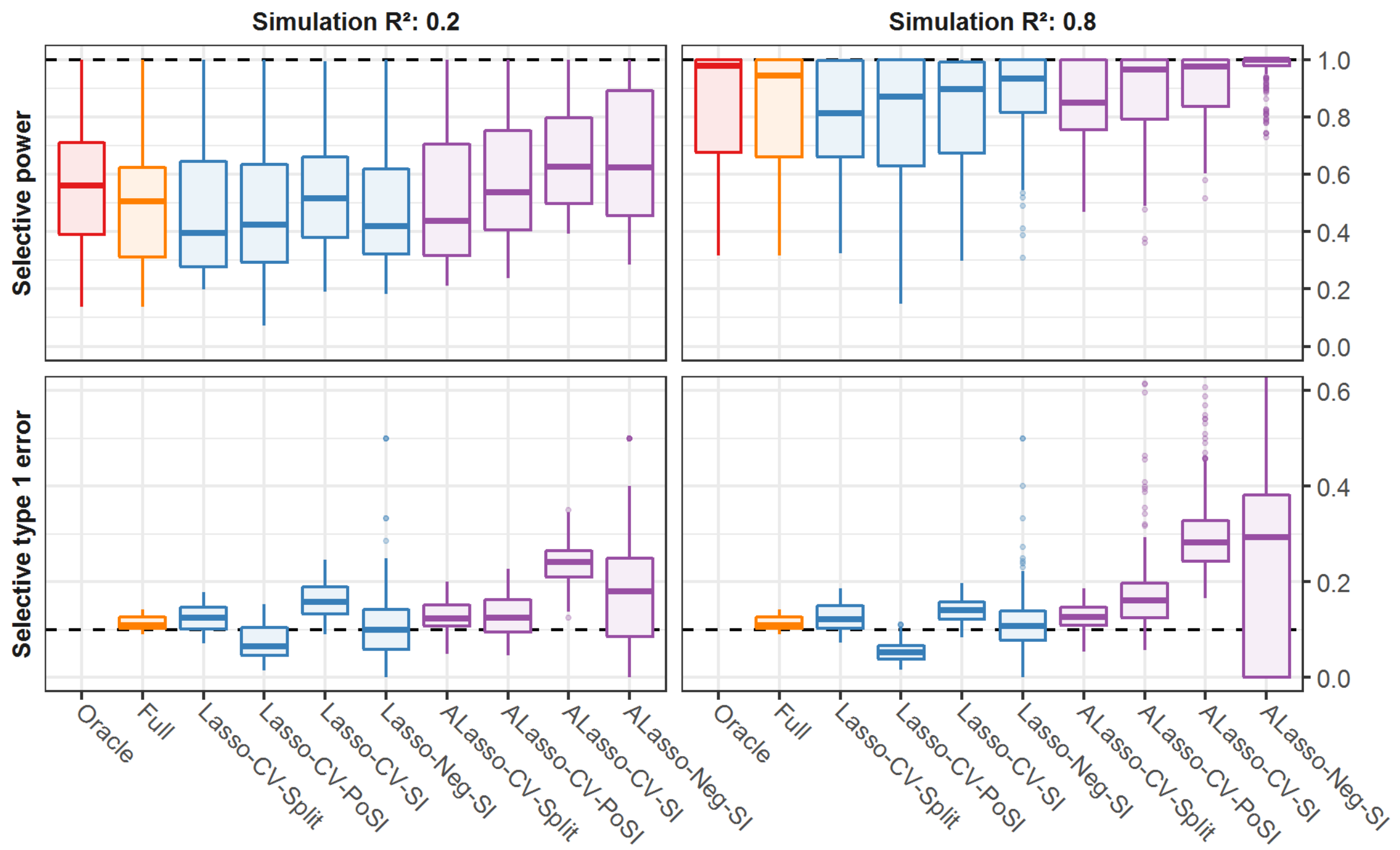

Figure 5: Summary of results from the toy simulation setup regarding the selective power (top row) and selective type 1 error (bottom row) of the selective 90% CIs for the submodel inference target (see Table 2). Results for all scenarios are summarised by boxplots. The target values are depicted as dashed lines (1 for power, the nominal significance level of 0.1 for type 1 error). Colors are based on the type of regression model used.

### Conditional primary estimands

In general, the results for the conditional primary estimands were very similar to the primary estimands, albeit with a higher variability due to the additional conditioning. Exemplary summary results for selective coverage conditional on inclusion of a specific variable of interest in both simulation setups are given in Supplementary Figure S 6. Noticeable were the increase in variability of selective coverage for the Lasso-Neg-SI and ALasso-Neg-SI methods, likely because some of the variables were selected very rarely in certain scenarios, leading to highly variable results. The Lasso-CV-Split and ALasso-CV-Split approaches showed decreased selective power for certain variables, due to efficiency loss by splitting the dataset. Figure 6 illustrates the association between selection frequency of a given variable of interest and selective coverage in the toy setup (for the realistic setup see Supplementary Figure S 7). Inference via the split-sample approach (Lasso-CV-Split and ALasso-CV-Split) or after Lasso without tuning (Lasso-Neg-SI and ALasso-Neg-SI) led to acceptable results, except for very low selection frequencies (i.e. below 20%). In the toy setup, the Lasso-CV-PoSI and ALasso-CV-PoSI approaches became increasingly unreliable with decreasing selection frequency. This is in-line with the results in Berk et al (3) regarding guarantees for conditional selective coverage. In the realistic setup this was no issue due to the larger number of variables and the resulting conservative CIs. The Lasso-CV-SI and especially the ALasso-CV-SI method also suffered from decreasing coverage with decreasing variable selection frequency. Variables markedly below 50% selection frequency led to low coverage rates in both simulation setups for the latter approach.

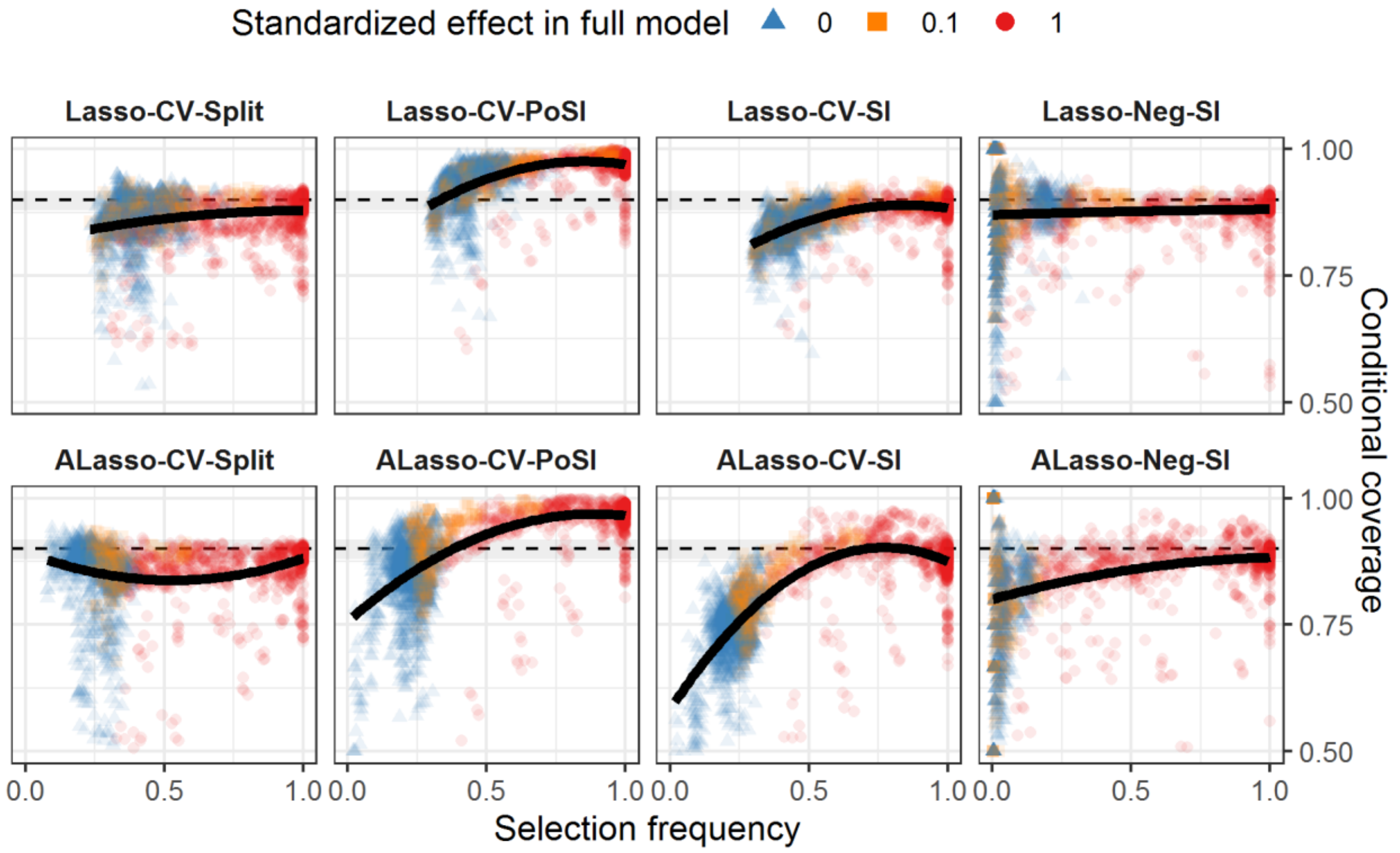

Figure 6: Comparison of selection frequency and conditional coverage from the toy simulation. Each dot represents results for a single variable in a specific simulation scenario. The target coverage value is depicted as dashed lines. Colors indicate if the variable is a predictor in the full model in the specific scenario. The black line provides a smoothed summary of the observed data (fitted with a quadratic B-spline term with 3 knots for selection frequency). To give an indication of variability expected in this simulation study, the grey areas indicate binomial 95% confidence intervals based on the number of iterations in each scenario (900) and the nominal confidence level. Ideally, the black line would be straight and on top of the dashed line at 0.9.

Widths of confidence intervals

Figure 7 and Supplementary Figure S 8 depict summaries of the results regarding the width of the selective CIs for the toy and realistic setup, respectively. Results are based on all simulation iterations and CI widths were standardized, corresponding to unit standard deviation of a predictor. The shortest intervals were obtained by the Oracle, followed by Full. The intervals obtained by the SI method (Lasso-CV-SI, ALasso-CV-SI, Lasso-Neg-SI, ALasso-Neg-SI) could be extremely wide and unstable, usually when a weak variable was selected into the active set. However, these CIs could be highly asymmetric, such that the power of the approaches to identify relevant predictors was on average not negatively impacted. For strong predictors, the CI widths got close to those of Full. In contrast, PoSI intervals (Lasso-CV-PoSI, ALasso-CV-PoSI) were always symmetric and shorter than the SI CIs. However, while their widths were quite consistent, that also meant that they were not able to narrow down in the case of strong predictors for which they tended to be the widest of all inference procedures. The ALasso-CV-SI and ALasso-Neg-SI intervals were generally shorter and less prone to extreme widths than the corresponding CIs for Lasso-CV-SI and Lasso-Neg-SI. A partial explanation might be ability of the adaptive Lasso to select relevant predictors more accurately. This trend became apparent in the realistic setup (see Supplementary Figure S 8). If runs with unstable inference were removed from the analysis, the width and variability of the selective CIs for the SI method decreased noticeably (not shown).

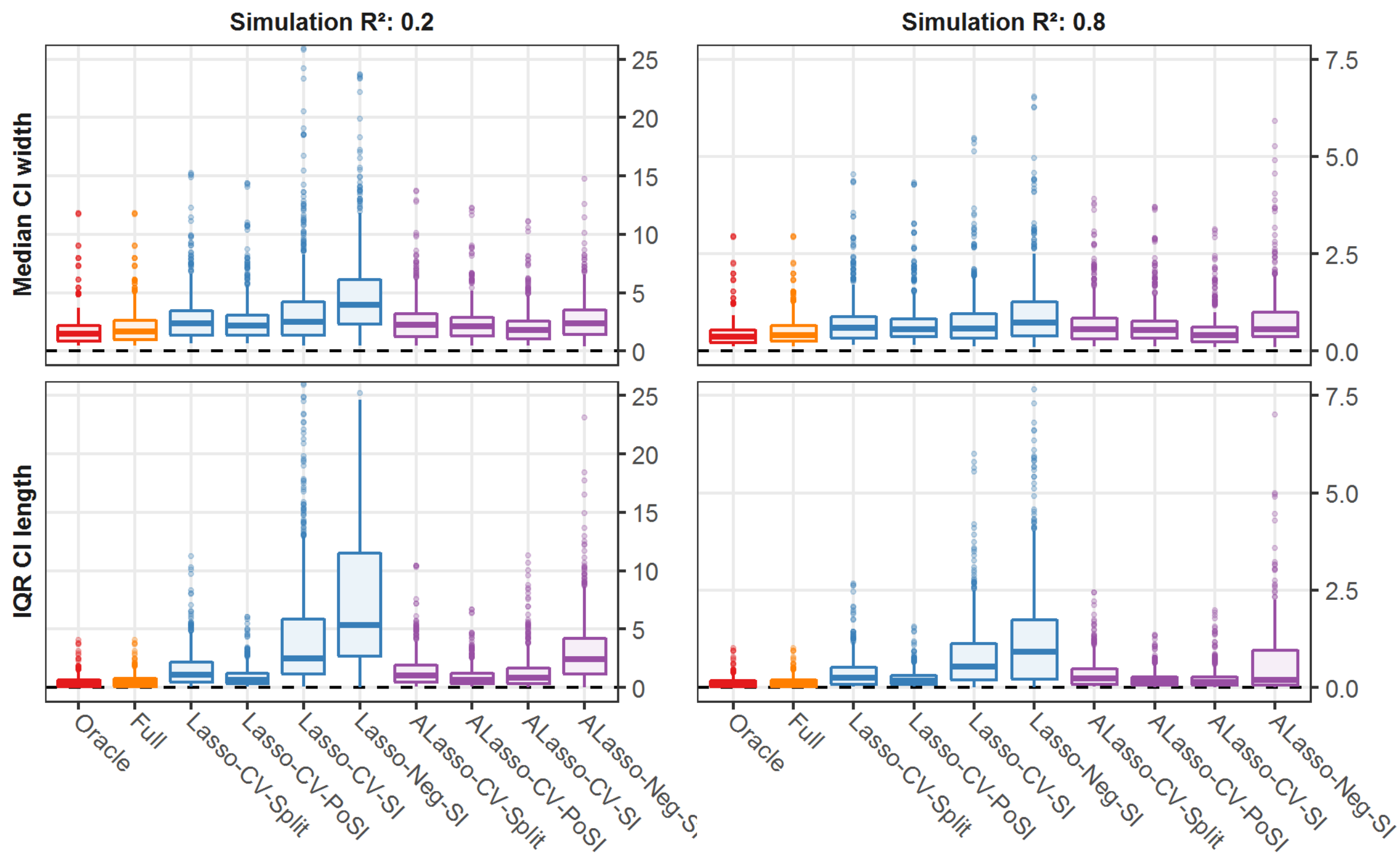

Figure 7: Summary of results from the toy simulation setup regarding the width (top row) and variability (bottom row) of the selective 90% CIs (standardized scale) for the submodel inference target (see Table 2). Results for all scenarios are averaged over all variables and are summarised by boxplots. Width zero is marked by dashed lines. Colors are based on the type of regression model used.

## 5.3. Predictive accuracy

Predictive accuracy on validation data behaved as expected and is depicted in Figure 8 and Supplementary Figure S 9 for the toy and realistic setups, respectively. With increasing sample size, the target $R^2$ values were achieved by the reference methods Full and Oracle, and most methods tuned by CV (Lasso-CV-SI, Lasso-CV-PoSI, ALasso-CV-SI, ALasso-CV-PoSI). However, Lasso-Neg-SI and ALasso-Neg-SI often led to very sparse models, resulting in inferior predictive accuracy. Similarly, for Lasso-CV-Split, ALasso-CV-Split only half the data was available to estimate effects and make predictions. This sub-optimal trade-off is clearly noticeable in the results.

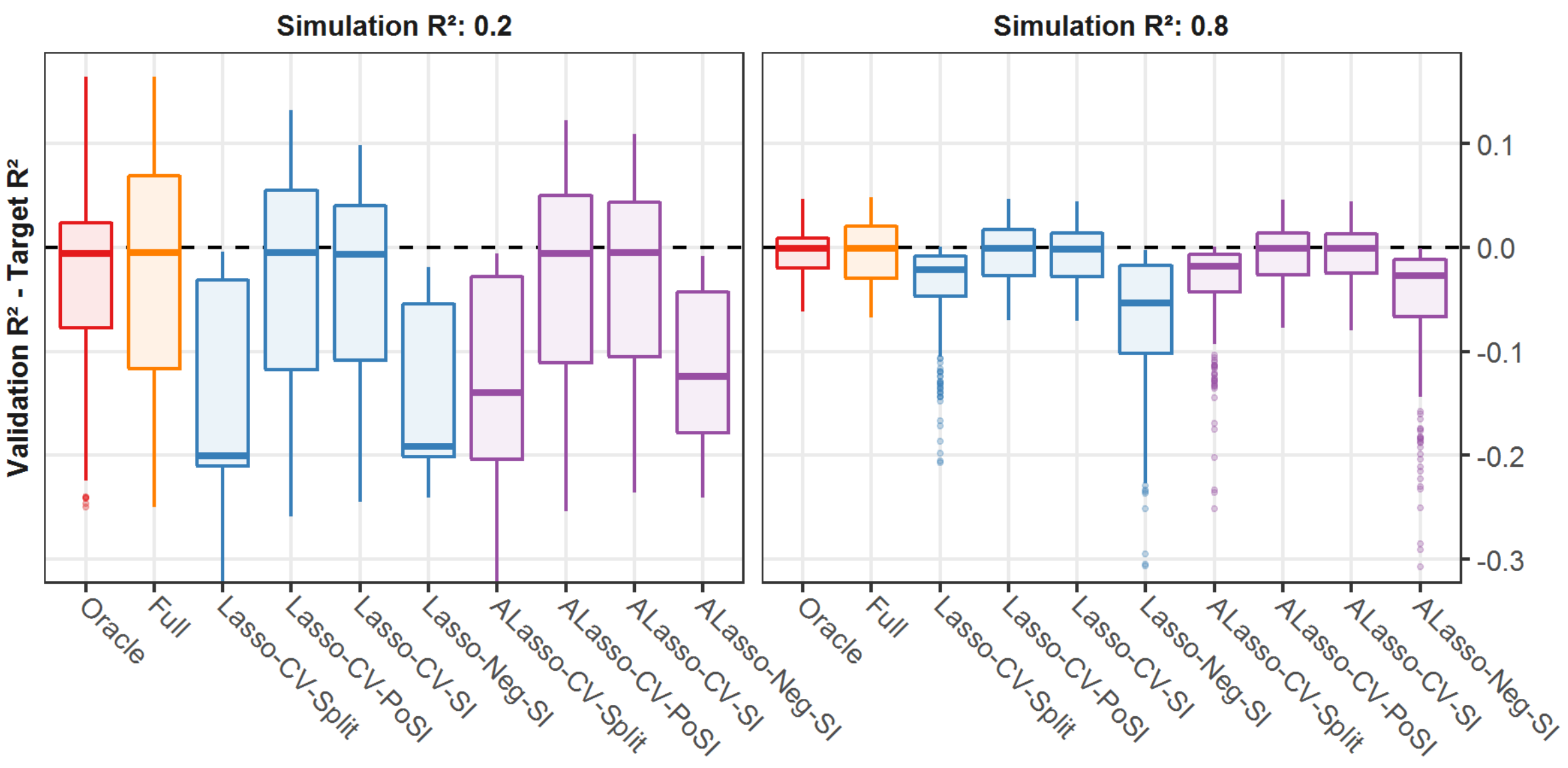

Figure 8: Summary of results from the toy simulation setup regarding predictive accuracy in terms of difference of validation $R^2$ and target $R^2$ (i.e. 0.2 in left panel, 0.8 in right panel). Results for all scenarios are summarised by boxplots. An optimal difference of zero is marked by dashed lines. Colors are based on the type of regression model used.

## 5.4. Other results

Computing time is an important aspect in practical applications, especially with many candidate predictors. As expected, the Lasso-CV-PoSI and ALasso-CV-PoSI approaches showed an exponential growth when comparing the toy and the realistic setup, while all of the other methods only showed minimal increases. The computation of the PoSI constant for a single iteration of the realistic setup with 17 candidate predictors took on average 35 seconds (Intel Core i7 4790 @ 3.6 GHz, R with multithreaded OpenBLAS matrix library). All the other methods remained well below 0.5 seconds.

## 5.5. Summary of simulation results

A high-level overview of the results is provided in Table 4. The Lasso-CV-Split and ALasso-CV-Split approaches led to generally acceptable estimates of coverage properties of the CIs even for weak predictors (i.e. within simulation variability for most scenarios), but this came at the cost of diminished predictive performance and statistical efficiency. The Lasso-CV-SI and ALasso-CV-SI approaches delivered acceptable results within expected simulation variability, but especially the latter could not guarantee nominal coverages. In particular, coverage was too low for variables with low selection frequencies (below 50%). This is in line with results in the literature but our results showed that the expected differences to nominal significance levels are small in situations with reasonable effect strengths. Lasso-Neg-SI and ALasso-Neg-SI led to slightly improved properties of the CIs, as well as lower probability to include weak predictors, but had a clear negative impact on prediction performance. The Lasso-CV-PoSI and ALasso-CV-PoSI methods led to extremely conservative confidence bounds and are probably mostly useful with a small to moderate number of candidate predictors of interest (e.g. a pool of 25 variables to be selected from).

**Conditional selective inference.** When a specific variable was of interest, ALasso-CV-SI had a higher-than-nominal conditional selective type 1 error, in contrast to ALasso-CV-Split and ALasso-CV-PoSI inference. A possible explanation might be that ALasso-CV-SI cannot properly account for the data-driven estimation of penalty weights, as it provides inference conditional on the model selected by the Lasso. The alternative approaches are agnostic to this issue; ALasso-CV-Split conditions on the whole dataset used for selection (thus also accounting for data-derived penalty weights) and ALasso-CV-PoSI provided valid inference regardless of the model selection procedure.

**Runtime and stability.** Two important take-aways from the study concern the runtime and the stability of inference computations, both of which are of practical importance to the applied researcher. While the Lasso-CV-PoSI and ALasso-CV-PoSI approaches led to an extremely low number of false selective significances, their application was computationally very demanding compared to all competing methods. Therefore they are likely most useful when only a few dozens of variables are to be submitted to selection. Berk et al gave some examples in which only part of the model space needs to be searched (section 4.5 (3)). On the other hand, the obtained confidence bounds were robust and had very low variance. In contrast, particularly the Lasso-CV-SI method suffered not only from highly variable confidence bounds, but also yielded unstable intervals excluding the coefficient point estimate in a non-negligible number of cases. The main reason for this was that the Lasso "just barely" included certain variables with a very small penalized coefficient, leading to issues in the computation of the CIs, as outlined in the original publication (30). A pre-specified (non-random) penalization parameter would not fix that issue. The target of submodel inference does not change by inclusion of noise variables in addition to true predictors, but their inclusion may result in extreme inflation of variance and numerical instability. Simply ignoring such unstable simulation runs led to improved selective coverage and lower selective type 1 error in our study, but in practice one cannot simply dismiss the data at hand and obtain another sample. Hence, stable inference requires a stable model in the first place. While this issue could be diagnosed in a practical analysis, the general solution is not evident. A pragmatic remedy (albeit introducing additional data dependent variability) would be to set a threshold value below which a penalized regression coefficient is considered to be equivalent to zero, therefore effectively removing a weak variable from the selected submodel. The selectiveInference package sets a default threshold for this. Furthermore, the package also returns one-sided p-values for each predictor in the submodel, assessing the individual null hypothesis whether the associated submodel coefficient is zero. These p-values seem to be more accurate than the associated CIs and led to plausible results, even in the case when the CIs were found to be unstable.

| Method | Stability | Coverage | Power | Type 1 error |
|---|---|---|---|---|
| Lasso_CV_Split | No concern | Acceptable | Low | Acceptable |
| Lasso_CV_PoSI | No concern | Too high | Low | Low |
| Lasso_CV_SI | Problematic | Acceptable | Acceptable | High |
| Lasso_Neg_SI | Problematic | Acceptable | Acceptable | Acceptable |
| ALasso_CV_Split | No concern | Acceptable | Acceptable | Acceptable |
| ALasso_CV_PoSI | No concern | Too high | Acceptable | Acceptable |
| ALasso_CV_SI | Problematic | Too low | High | High |
| ALasso_Neg_SI | Problematic | Acceptable | High | High |

Table 4: Overview of main results for the primary estimands of our simulation study. By "Acceptable" we mean that results were mostly (i.e. in median over all scenarios) within the expected simulation variability.

## 6. Real data example

We use Johnson's body fat dataset (54) as a real data example to demonstrate the practical application of the selective inference framework. The underlying research question was to estimate the percentage of body fat in 252 men, measured by underwater weighting according to Siri's formula (55), using multiple linear regression. One individual, also reported in Johnson's original publication, was removed from the analysis due to implausible values. The candidate predictors are age (in decades), height (dm), weight (kg) and ten anthropometric body measurements (all in cm). Some variable units were converted for better visualisation (age from years to decades, height from cm to dm). The dataset is freely available from the original articles' website (56). An interesting feature of the dataset is its correlation structure: the ten body measurements and weight are highly correlated (mean Pearson correlation of 0.65 between the individual variables), while age and height are rather uncorrelated (mean Pearson correlation of 0.03 and 0.28 to all other variables, respectively). It is therefore similar to the block correlation design we have used in our simulation setups.

**Research question.** The goal of variable selection in this case study is to optimize the number of measurements necessary for the body fat estimation in future applications, rather than studying their causal relationship with the outcome. In such situations the submodel view of post-selection inference is appropriate. Note that due to the correlation structure of the dataset, even variables excluded from the finally selected model cannot necessarily be deemed as "not predictive", since they are very likely to be correlated to a "predictor" in the final (sub)model. It is therefore natural to be interested in inference about the specific set of active variables, rather than targeting the full model. In the latter case, any assumption about the correctness of the chosen submodel would be questionable due to the high correlations.

**Analysis.** We analysed the dataset with the methods of our study (Table 3). As observed in our simulations (see Section 5.5), and in-line with recent recommendations on practical application of variable selection (2), we found it beneficial for the interpretation to additionally investigate variable selection frequencies. We computed these using 100 bootstrap resamples of the dataset. We give an overview of the results in Figure 9.

**Results.** Variable selection frequencies of the selected variables were mostly above 50%, and for variables with lower frequencies the CIs had reasonable widths; therefore selective inference should be generally reliable. Figure 9 provides a comparative presentation of the results, but in a real analysis each model would be interpreted by itself. As an example, assume it was decided to use the Lasso tuned by CV for variable selection, followed by SI for selective inference (Lasso-CV-SI). This procedure selects 4 variables (abdomen, wrist, age and height) with frequency close to 100%, and their 90% CIs exclude zero. Judging by their standardized coefficient estimates and selective CIs, the importance of the other variables as predictors is less certain, even if they were selected into the final model with high bootstrap selection frequencies. Therefore, this example shows that the selective CIs are useful to assess the relative importance of predictors, and to judge which predictors are possibly exchangeable in future applications. This is especially useful when the Lasso is used for variable selection, as this method tends to include too many predictors. The example also illustrates that selective CIs can be very asymmetric. In comparison, the Lasso-CV-PoSI approach leads to wider CIs and can only confirm the importance of abdomen and wrist as strong predictors. Lasso-CV-Split leads to a smaller model and three significant variables (abdomen, height, wrist), but with similar point estimates as the former two approaches. These different properties express different use-cases: the SI method could be potentially useful in early phase, explorative stages to single out weak predictors surviving the Lasso screening; the PoSI method could be favoured in later phase research where false positives are of particular concern and inference should respect all kinds of selection mechanisms during analysis. In this example, splitting the dataset into two halves leads to a ratio of observations per variable below 10. In comparable scenarios of our simulation study (see

Section 5.5), this method often had smaller predictive accuracy than the former two approaches. This is even more likely for the Lasso-Neg-SI method, which leads to an extremely small model, containing only two variables.

If the ALasso was chosen for variable selection, the resulting models are generally sparser than for the Lasso, but in this example the resulting point and interval estimates are similar given the relatively high variable selection frequencies (mostly above 50%). Thus, the ALasso might be an attractive alternative in this data example.

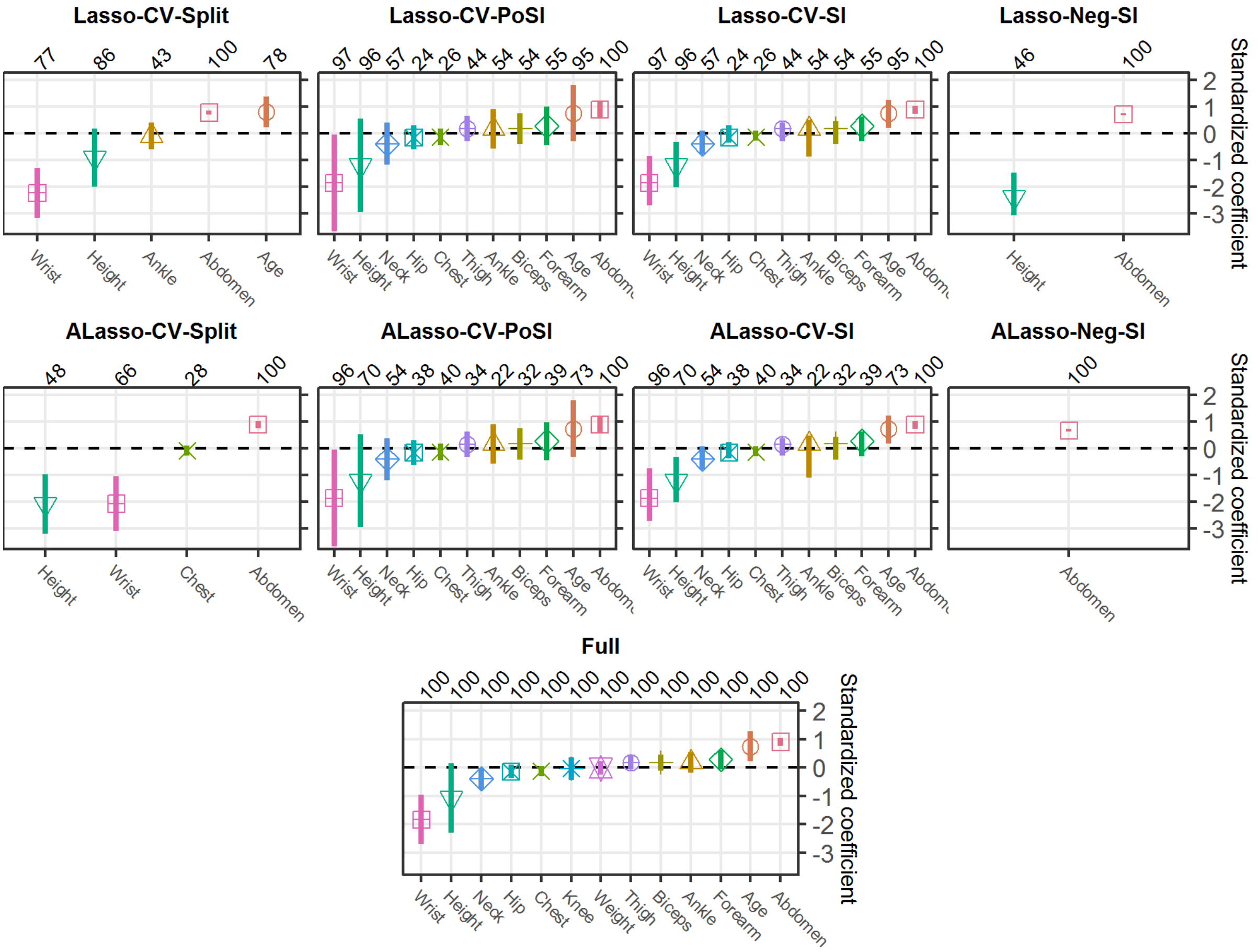

Figure 9: Overview of the results for the body fat dataset (shown for original scales of the variables). The panels depict point estimates and 90% selective CIs from the different methods. The coefficients are ordered from left to right by increasing standardized coefficients. The individual selection frequencies estimated by 100 bootstrap resamples are given by as percentages above each panel.

## 7. Discussion

The problem of selectively assessing hypotheses has been recognized for a long time already in statistical research, and identified as partly responsible for the lack of replicability of scientific research (3, 57). The formalisation of the selective inference framework, as well as the submodel view of variable selection, are rather recent developments, re-vitalizing the research of this long-standing issue. With this work, we attempted to bridge the gap between theoretical understanding via asymptotic results, and practical application of novel methodology in actual, finite sample. In particular, selective inference requires careful consideration of the intended use of a statistical model by the practitioner. The interpretation of inference for hypotheses which are explored and addressed using the same set of data is not as clear as in the classical framework, in which inference addresses an a-priori fixed hypothesis. However, in routine work on descriptive or predictive research problems the interest often lies in the finally selected predictors, rather than addressing the population parameters of an underlying data generating mechanism. Similarly in explanatory research, the data-adaptive selection of confounders may provide benefits when estimating the exposure effects, given all the causal assumptions remain intact or are addressed by specialized methods. Selective inference is well matched to the philosophy of each of these usage cases, where it is not of primary concern whether a "true model" exists and what its parameters could be. Through accounting for the selection of the model of interest, over-optimism in reported inference can be reduced, thereby enhancing replicability, and allowing to make informed decisions about the future use of the model's parameters. Selective inference in such a setting therefore seems natural and makes the research goal more explicit.

**Practical role of selective inference.** In our view, selective inference addresses the detection of weak predictors and correcting for over-optimism after selection. This was elaborated in our real data example, where an additional inference step after selection potentially allows to single out those variables which are likely most relevant to the prediction of the outcome, in contrast to those with small, but non-zero point estimates. This is of great interest when using the Lasso, as it is known to have a high probability for the inclusion of weak predictors.

However, selective inference does not come for free, as there is not only a trade-off regarding power and type 1 error, but also a relation of selection quality and inference accuracy. For example, as we have seen in our simulation study, the SI approach to selective inference could only provide reliable results when the selected submodel was stable in the first place (i.e. only included variables with comparatively high selection frequency).

**Limitations.** Our study was restricted to recent approaches to selective inference, but this allowed to us to focus on the underlying framework based on the submodel view of inference, which we believe requires careful interpretation in practical applications. We have reduced the variable selection problem to the bare minimum to explore the main differences between the methods in our study without the distractions when the parameter space is large and unintelligible. The data generating mechanism of our simulations was chosen to be sparse and all effects were assumed to be linear, fitting to the assumptions that underlies the Lasso model selection procedures. We used mostly default parameters for the implementation of the methods as long as they provided sensible results, ensuring comparability between the methods. Lastly, for the PoSI and SI methods we derived an estimate for the outcome variance from the residual variance of the full model, which is in-line with the literature. This reflects that is often difficult to obtain an independent assessment of the outcome variance in practice, as assumed by PoSI and SI. However, given that the full model generally contained weak variables in our simulation setup, this likely underestimated the true outcome variance.

**Scope of the results.** The results in this manuscript were reported for 90% selective CIs, but similar relative performances of the inference methods are expected for other coverage levels. This was corroborated by a smaller simulation study for 95% selective CIs.

In this paper we have focused on the use of CV to determine the penalization strength of the Lasso regression models. This method was chosen due to its widespread use and as it is in close relationship with another commonly used tuning criterion, the Akaike information criterion (AIC). Both tune penalization strength by an estimate of the out-of-sample prediction error based on the observed outcomes. While the resulting penalization parameters differ, we found in a smaller simulation based case-study that frequentist properties of selective CIs between CV and AIC were comparable. Future work will provide a more in-depth comparison.

We expect that our results approximately hold for other types of regression models as well, such as logistic regression. This was also assessed in an explorative smaller simulation study, in which the essential conclusions were essentially similar to those presented here. However, the PoSI method is so far only available for linear regression and could not be evaluated for logistic regression.

**Final remarks.** In general we recommend to combine post-selection inference for predictive or descriptive research questions with an assessment of model stability and variable selection frequencies. We found the SI methodology for selective inference to be quite promising with acceptable coverage results in most scenarios. It is therefore our recommended approach in most cases. If the number of observations is large and simplicity is favoured over efficiency, then the Split approach to selective inference, or possibly a more refined variant (data carving (8, 28)), could be a viable alternative. If only few predictors are submitted to variable selection and the primary concern is to avoid false positive significances, then the conservative and robust PoSI approach could be considered. For the adaptive Lasso, especially if the penalization strength is tuned, only the Split and PoSI approaches are recommended as they were able to uphold coverage guarantees, but the SI method is problematic for weak predictors. Concluding, we found selective inference to be a potentially useful inferential tool for predictive and descriptive regression models using the Lasso. However, the practical application requires sophisticated interpretation and awareness of the intended use of the model and some dedication while working with currently not-yet user friendly software packages.

# Acknowledgments

We thank the members of the prognosis research group at the Section for Clinical Biometrics of the Medical University of Vienna, Harald Binder, Willi Sauerbrei and Kathrin Blagec for their comments throughout this project.

# Supplementary material

## Evaluating methods for Lasso selective inference in biomedical research by a comparative simulation study

Michael Kammer[1,2], Daniela Dunkler[1], Stefan Michiels[3] and Georg Heinze[1]

1 Medical University of Vienna, Center for Medical Statistics, Informatics and Intelligent Systems, Section for Clinical Biometrics, Vienna, Austria

2 Medical University of Vienna, Department for Internal Medicine III, Division of Nephrology and Dialysis, Vienna, Austria

3 Service de Biostatistique et d'Epidémiologie, Gustave Roussy; INSERM, CESP U1018, University Paris-Saclay, Villejuif, France

## 1. Usage of selectiveInference package

The selectiveInference package for R was found to have some minor bugs during this simulation study which the applied researcher should be aware of. They concern logistic regression. These have also been reported as issues on the package's Github repository.

- Confidence intervals for logistic regression may have the wrong sign. After studying the code, we believe this happens when the sign of the coefficient is negative, in which case the interval needs to be flipped.
- The passed significance level for logistic regression is ignored. A simple fix for this was done in the software used for this simulation study.

# 2. Simulation profile

This brief overview provides a high-level summary and pointers to the essential definitions of the simulation study in this manuscript.

## 2.1. Design

*Aim:* Evaluate several recent proposals to conduct inference after variable selection in the context of Lasso regression regarding their frequentist properties in practical usage scenarios.

*Data generation:* Aimed at studying different correlation patterns and effect distribution in sparse settings. Data generation procedure outlined in the figure below.

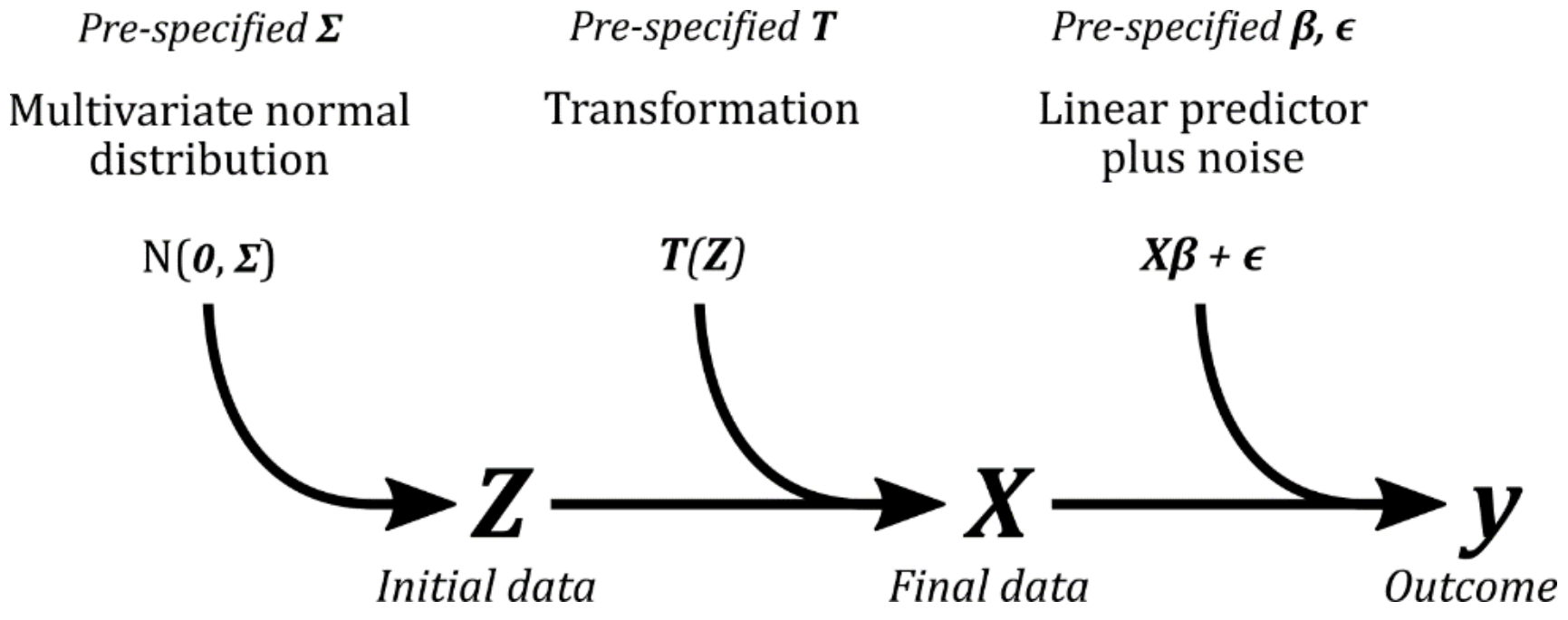

Two setups with different pre-specifications (notation referencing figure above):

| | | **Toy setup** | **Realistic setup** |
|---|---|---|---|
| **Fixed design parameters** | Motivation | Simplicity, insight | Realistic data |
| | Regression type | Gaussian | Gaussian |
| | Number of variables | 4 | 17 |
| | Type of variables | Continuous | Continuous, binary |
| | Distribution of variables | Gaussian | Mixed (specified in Supplementary Table S 2) |
| **Varying design parameters** | Correlation structures $\boldsymbol{\Sigma}$ | 7 blocked correlation matrices with no or strong correlation | Fixed, mimicking real study (Supplementary Figure S 2) |
| | Coefficient structures $\boldsymbol{\beta}$ | 10 (Supplementary Table S 1) | 13 (Supplementary Table S 3) |
| | True target $R^2$ (noise $\boldsymbol{\epsilon}$) | 0.2, 0.5, 0.8 | 0.2, 0.5, 0.8 |
| | Observations per variable | 5, 10, 50 | 5, 10, 50 |
| **Simulation parameters** | Number of scenarios | 630 | 117 |
| | Iterations per scenario | 900 | 900 |

*Primary estimands:* selective coverage, power and type 1 error (Table 2)
*Conditional primary estimands:* selective coverage, power and type 1 error conditioning on the event that a specific variable of interest is selected (Table 2)
*Further estimands:* true model and variable selection frequencies, width of confidence intervals, prediction accuracy, inference stability.

*Methods:* Lasso and adaptive Lasso for variable selection, tuned with either CV or using a fixed estimated penalization strength following Negahban et al (1). Sample splitting, or the approaches by Berk et al (2) and Lee et al (3) for selective inference. See Table 3.

*Performance measures:* Difference to nominal significance level for selective coverage and type 1 error. Highest value for selective power. Minimal median and interquartile range per scenario for confidence intervals. Validation $R^2$ for prediction accuracy.

## 2.2. Coding and execution

Study conducted in R version 3.5.1. using the packages glmnet (version 2.0-18) (4), selectiveInference (version 1.2.4) (5) and PoSI (version 1.0) (6). Data was generated using the simdata package (version 0.5.0.9000) (7).

## 2.3. Analysis

*True model and variable selection:* Adaptive Lasso more often selects true data generating model and less often includes false positives than Lasso (Figure 3, Supplementary Figure S 3, Supplementary Figure S 4).

*Stability of inference:* Method by Lee et al after CV tuned Lasso provided confidence intervals not including the point estimate or having infinite width in a non-negligible fraction of iterations (30% for real setup, 5% for toy setup). Issue less severe for adaptive Lasso (7% and 1% respectively).

*Primary estimands:* Selective coverage and type 1 error mostly acceptable for all inference approaches for Lasso, undercoverage for adaptive Lasso tuned by CV for the Lee et al method; selective power generally lower for Berk et al approach but very low type 1 error (Figure 4, Figure 5, Supplementary Figure S 5).

*Conditional selective coverage:* Inference for a fixed variable of interest dependent on the variable's selection frequency (undercoverage for rarely selected variables), critical in case of adaptive Lasso tuned by CV and the Lee et al method (Figure 7, Figure 6, Supplementary Figure S 7).

*Width of confidence intervals:* Highly variable and generally narrower for adaptive Lasso than for Lasso for Lee et al method (Figure 7, Supplementary Figure S 8).

*Prediction performance:* Methods using split-sample and a fixed estimated penalization parameter had low prediction performance (Figure 8, Supplementary Figure S 9).

# 3. Simulation study details

## 3.1. Derivation of estimators

We present the derivation of our simulation approximation of the selective estimands exemplary for the case of selective coverage after certain variable selection procedure of interest. To begin with, let the model $M \subseteq M_F = \{1, \dots, p\}$ be fixed and let $i \in M$. We denote the complete set of simulation iterations as $S = \{1, \dots, n_{sim}\}$ for a fixed total number of simulations $n_{sim}$. As in the main manuscript, we use the notation $\widehat{M}$ to indicate the random variable representing the model selected by the variable selection procedure. By the use of $\widehat{M}_s$ we denote the model chosen in a specific iteration $s$ of the simulation study.

Then, in each iteration $s$ of our simulation study in which the selected model $\widehat{M}_s$ coincides with $M$, we observe the event if a selective CI for $i$ covers its target parameter or not (if $\widehat{M}_s \neq M$ inference is neither available, nor of interest to us). From this we can estimate the conditional coverage probabilities $\mathbb{P}\left[\beta_{i,M} \in CI_{i,M} \middle| \widehat{M} = M, i \in \widehat{M}\right] = \frac{\sum_{s \in S} \mathbb{I}[\widehat{M}_s = M \wedge \beta_{i,M} \in CI_{i,M}]}{\sum_{s \in S} \mathbb{I}[\widehat{M}_s = M]}$, where we use $\mathbb{I}$ to denote the indicator function. The formula can be interpreted as the number of all iterations where the CI covered its target parameter (provided it exists), divided by the frequency how often the specific model $M$ was selected. If $M$ was never selected, then the conditional coverage probability is estimated as zero, but it plays no role in the further computations. By the law of total probability we can use these probabilities to compute the conditional coverage probabilities for a fixed variable $j \in M_F$:

$$\mathbb{P}\left[\beta_{j,\widehat{M}} \in CI_{j,\widehat{M}} \middle| j \in \widehat{M}\right] = \sum_{M \subseteq M_F} \mathbb{P}\left[\beta_{j,M} \in CI_{j,M} \middle| \widehat{M} = M, j \in \widehat{M}\right] \mathbb{P}\left[\widehat{M} = M \middle| j \in \widehat{M}\right].$$

The latter term $\mathbb{P}\left[\widehat{M} = M \middle| j \in \widehat{M}\right]$ for fixed $M$ and $j$ is estimated by $\frac{\sum_{s \in S} \mathbb{I}[\widehat{M}_s = M]}{\sum_{s \in S} \mathbb{I}[j \in \widehat{M}_s]}$. Note that for models $M$ which do not contain the variable of interest, $\mathbb{P}\left[\widehat{M} = M \middle| j \in \widehat{M}\right] = 0$ (i.e. in such cases inference is not available).

For the overall expected selective coverage we further marginalize over all candidate predictors.

$$\mathbb{P}\left[\beta_{.,\widehat{M}} \in CI_{.,\widehat{M}}\right] = \sum_{j \in M_F} \mathbb{P}\left[\beta_{j,\widehat{M}} \in CI_{j,M} \middle| j \in \widehat{M}\right] \mathbb{P}\left[j \in \widehat{M}\right].$$

Note that the term $\mathbb{P}\left[j \in \widehat{M}\right]$ can be derived from the variable selection frequencies $\frac{\sum_{s \in S} \mathbb{I}[j \in \widehat{M}_s]}{n_{sim}}$ through re-normalisation by $1/n_{sim} \sum_{k \in M_F} \sum_{s \in S} \mathbb{I}\left[k \in \widehat{M}_s\right]$ such that the result is a probability distribution over the candidate predictors, i.e. $\sum_{j \in M_F} \mathbb{P}\left[j \in \widehat{M}\right] = 1$. The expression $\sum_{k \in M_F} \sum_{s \in S} \mathbb{I}\left[k \in \widehat{M}_s\right]$ is simply counting the total number of selection events for the variable selection procedure over all iterations of the simulation scenario.

In practice the computations simplify drastically, resulting in the estimators shown in Table 2. For example, the approximation of overall selective coverage can be explicitly obtained from the equations above as

$$\mathbb{P}\left[\beta_{.,\widehat{M}} \in CI_{.,\widehat{M}}\right] = \sum_{j \in M_F} \sum_{M \subseteq M_F} \frac{\sum_{s \in S} \mathbb{I}\left[\widehat{M}_s = M \wedge \beta_{j,M} \in CI_{j,M}\right]}{\sum_{s \in S} \mathbb{I}[\widehat{M}_s = M]} \cdot \frac{\sum_{s \in S} \mathbb{I}[\widehat{M}_s = M]}{\sum_{s \in S} \mathbb{I}\left[j \in \widehat{M}_s\right]} \cdot$$

$$\frac{\sum_{s \in S} \mathbb{I}\left[j \in \widehat{M}_s\right]}{n_{sim}} \cdot \frac{n_{sim}}{\sum_{k \in M_F} \sum_{s \in S} \mathbb{I}\left[k \in \widehat{M}_s\right]},$$

which allows significant simplifications. Note that the equation on the right hand can then be easily interpreted as “evaluate whether a CI covers its target parameter, whenever a CI is available". In the case of a conditional quantity, the analogous interpretation would be "evaluate whether a CI for variable $j$ covers its target parameter, whenever a CI for $j$ is available (i.e. $j$ was selected)". Similar interpretations apply to the other selective quantities in Table 2.

## 3.2. Toy simulation setup

The toy setup was kept extremely simple with four standard normal distributed candidate predictors in order to provide a focused assessment of the methods in our study. The rationale was that more variables just meant more (hard to control) noise and a larger number of essentially equivalent ways to distribute effects, blurring the conclusions and making the results less intelligible. Similar, explorative simulation setups using a larger number of variables were conducted to corroborate the findings from this small setup. Using this distilled setup still allowed us to study a variety of 7 different structures for the true correlation matrix:

- Uncorrelated (1 matrix design): all variables were uncorrelated
- Correlated (2 matrix designs): all variables were equally correlated (either 0.8 or -0.8).
- Two 2x2 blocks (2 matrix designs): the correlation matrix consisted of two blocks of size 2x2. Correlation within the blocks was either the same (0.8) or mixed (0.8 / -0.8 respectively). There was no correlation between the blocks of variables.
- One 3x3 block (2 matrix designs): the correlation matrix consisted of two blocks, one of size 1x1 (i.e. a single variable) and one of size 3x3. The last 3 variables were equally correlated (either 0.8 or -0.8). There was no correlation between the blocks of variables.

All used correlation values were chosen rather high in order to have a strong impact on the results. We defined 10 possible structures for the vector of true regression coefficients $\boldsymbol{\beta} = (\beta_1, \beta_2, \beta_3, \beta_4)$, with (standardized) effect strengths of either 1 or 0.1. These are listed in Supplementary Table S 1Supplementary Table S 1: Overview of coefficient structures in toy simulation setup.. We used a full factorial design, resulting in a total of 630 scenarios for this toy setup. Each individual simulation scenario comprised 900 repetitions. Example code demonstrating how the data was simulated using the simdata package (7) is provided in the file *Paper_Toy_Setup_Demo.R*.

| Number | Coefficients | Remarks |
|---|---|---|
| 1 | $\beta_1 = 1$ | |
| 2 | $\beta_1 = \beta_2 = 1$ | |
| 3 | $\beta_1 = 1, \beta_2 = 0.1$ | strongly differential effect size, Lasso is known to have difficulty dealing with such situations |
| 4 | $\beta_1 = \beta_2 = \beta_3 = \beta_4 = 1$ | non-sparse situation |
| 5 | $\beta_1 = \beta_3 = 1$ | differences to 2. due to block correlation structure |
| 6 | $\beta_1 = 1, \beta_3 = 0.1$ | differences to 3. due to block correlation structure |
| 7 | $\beta_1 = 0.1, \beta_3 = 1$ | differences to 3. And 6. due to block correlation structure |
| 8 | $\beta_3 = 1$ | differences to 1. due to block correlation structure |
| 9 | $\beta_3 = \beta_4 = 1$ | differences to 2. and 5. due to block correlation structure |
| 10 | $\beta_3 = 1, \beta_4 = 0.1$ | differences to 6. due to block correlation structure |

Supplementary Table S 1: Overview of coefficient structures in toy simulation setup. Only non-zero coefficients are specified on a standardized scale.

## 3.3. Realistic simulation setup

The realistic setup featured a fixed, complex correlation structure and different kinds of variable distributions based on real clinical data as presented in (8). After drawing standard normal data using a pre-specified correlation matrix (see Supplementary Figure S 1), the final simulated dataset was obtained by applying transformations according to Supplementary Table S 2. The 17 final predictors comprise continuous and discrete variables, with clusters of highly correlated variables as well as uncorrelated ones (multiple correlation coefficients between 0 and 0.7). We considered 13 different options for the vector of standardized regression coefficients = $(\beta_i)_{i=1,\dots,17}$ , based on the correlation network to provide the simulation with interesting correlation scenarios. The values are listed in Supplementary Table S 3. In this setup a full factorial design led to 117 scenarios.

This setup was based on the design presented in (8). Example code demonstrating how the data was simulated using the simdata package (7) is provided in the file *Paper_Real_Setup_Demo.R*.

*Correlation structure*

The correlation matrix used for drawing data from a multivariate normal distribution was modified from the original to feature stronger correlations. The following network depicts the final correlation structure. Individual variables as nodes in a graph, correlations between two variables are indicated as an edge with specified correlation coefficient.

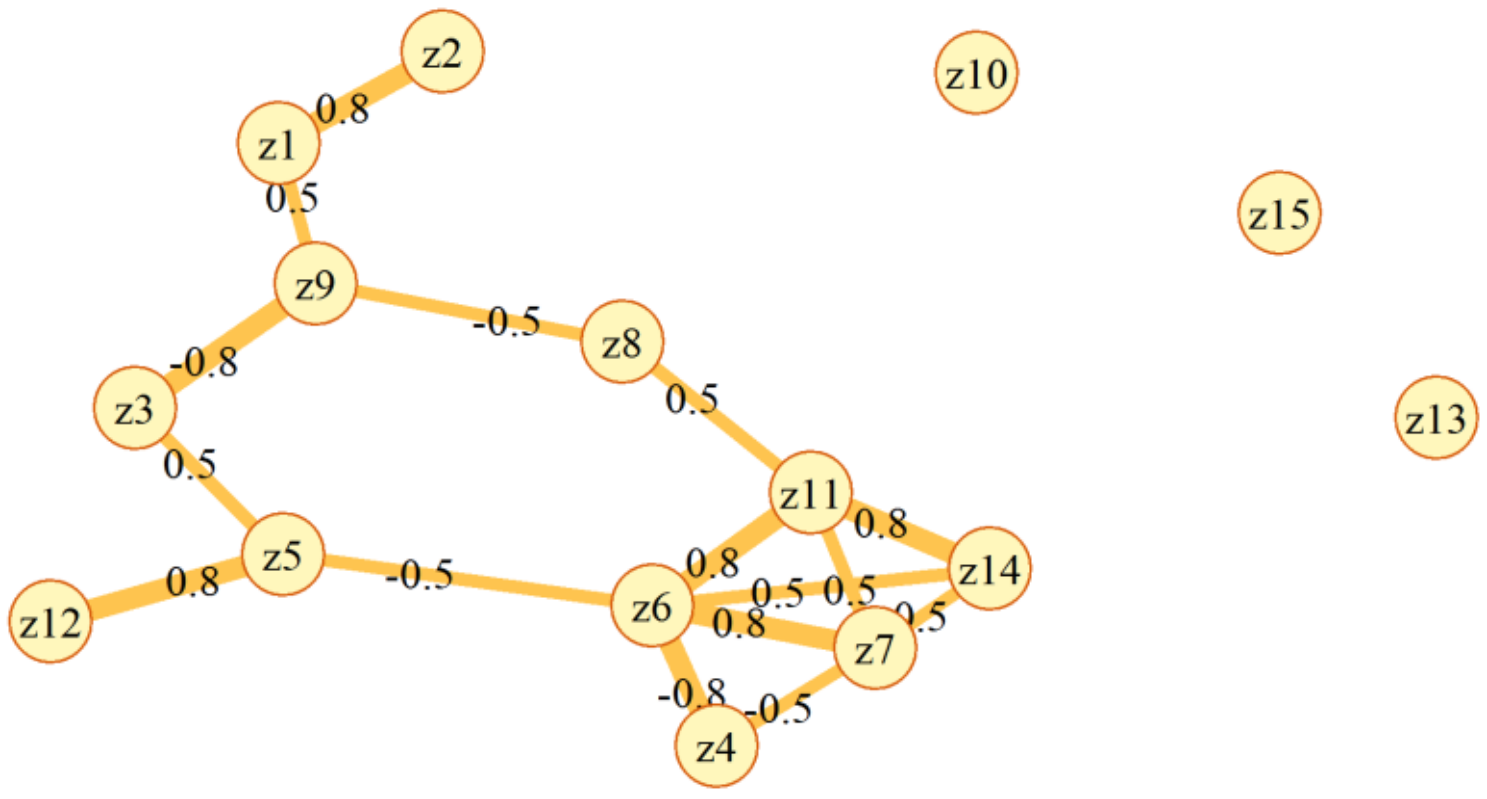

Supplementary Figure S 1: Initial correlation network for real simulation setup.

*Data transformation*

Data from the initial multivariate distribution was transformed to achieve different variable distributions. The transformations were the same as in the original publication:

| Initial variable | Final variable | Type | Multiple correlation |
|---|---|---|---|
| $z_1$ | $v_1 = [10z_1 + 55]$ | Continuous | 0.63 |
| $z_2$ | $v_2 = I(z_2 < 0.6)$ | Binary | 0.58 |
| $z_3$ | $v_3 = \exp(0.4z_3 + 3)$ | Continuous, skewed | 0.52 |
| $z_4$ | $v_4 = I(z_4 \geq -1.2)$ | Ordinal | 0.44 |
| $z_4$ | $v_5 = I(z_4 \geq 0.75)$ | Ordinal | 0.35 |
| $z_5$ | $v_6 = \exp(0.5z_5 + 1.5)$ | Continuous, skewed | 0.68 |
| $z_6$ | $v_7 = [\max(0, 100\exp(z_6) - 20)]$ | Continuous, skewed | 0.66 |
| $z_7$ | $v_8 = [\max(0, 80\exp(z_7) - 20)]$ | Continuous, skewed | 0.65 |
| $z_8$ | $v_9 = I(z_8 < -0.35)$ | Binary | 0.43 |
| $z_9$ | $v_{10} = I(0.5 \leq z_9 < 1.5)$ | Ordinal | 0.46 |
| $z_9$ | $v_{11} = I(1.5 \leq z_9)$ | Ordinal | 0.47 |
| $z_{10}$ | $v_{12} = 0.01[100(z_{10} + 4)^2]$ | Continuous | 0.01 |
| $z_{11}$ | $v_{13} = [10z_{11} + 55]$ | Continuous | 0.71 |
| $z_{12}$ | $v_{14} = [10z_{12} + 55]$ | Continuous | 0.63 |
| $z_{13}$ | $v_{15} = [10z_{13} + 55]$ | Continuous | 0.01 |
| $z_{14}$ | $v_{16} = I(z_{14} < 0)$ | Binary | 0.61 |
| $z_{15}$ | $v_{17} = I(z_{15}10)$ | Binary | 0.01 |

Supplementary Table S 2: Final variables for real simulation setup. Multiple correlation was estimated from a simulated dataset with 100000 observations.

The final correlation network was similar to the original one.

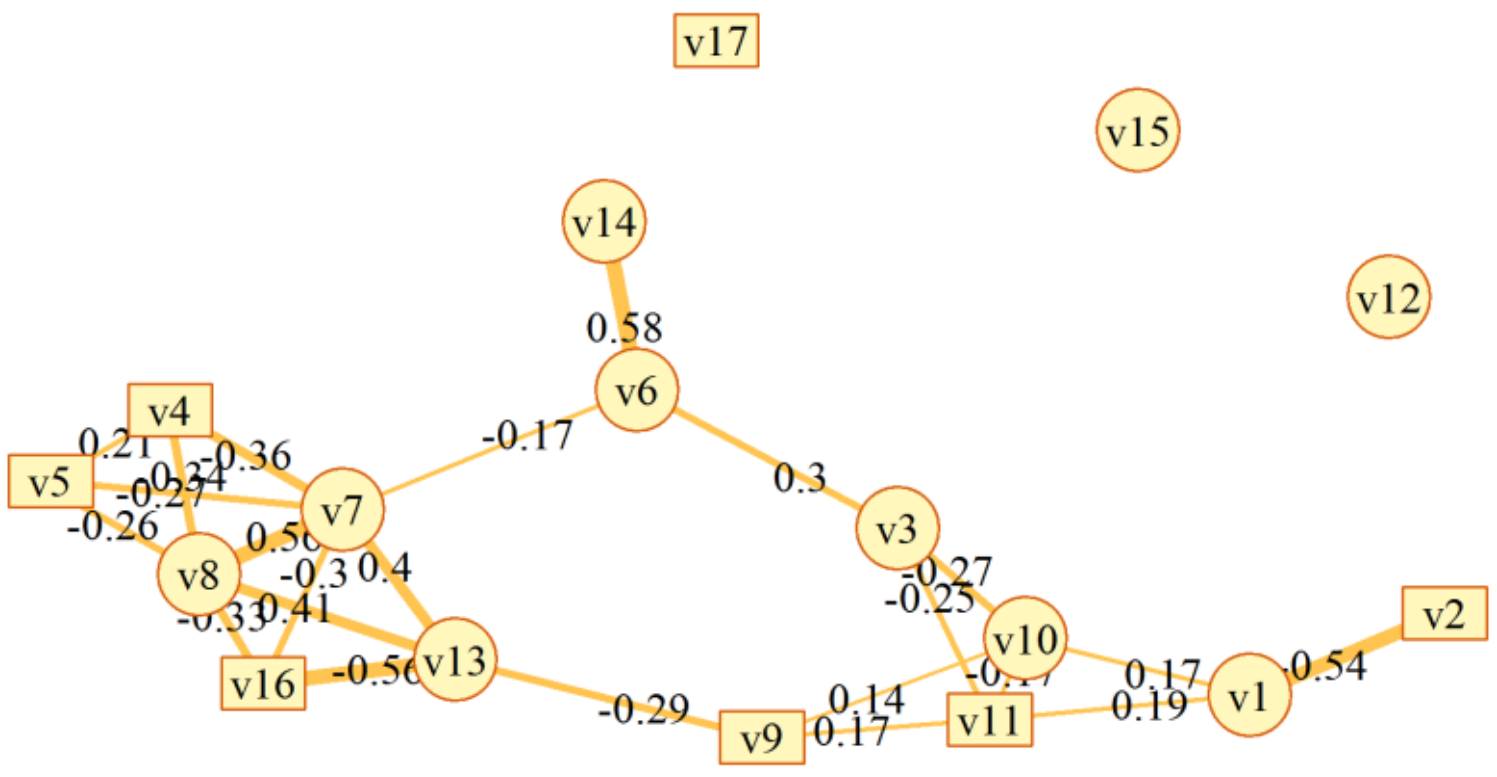

Supplementary Figure S 2: Final correlation network for real simulation setup. Discrete (binary, ordinal) variables are depicted as squares, continuous ones as circles. Correlations were estimated from a simulated dataset with 100000 observations.

*Coefficients*

| Number | Coefficients | Remarks |
|---|---|---|
| 1 | $\beta_2 = \beta_4 = \beta_{14} = 1$ | 3 rather independent variables far apart in the network |
| 2 | $\beta_7 = \beta_8 = \beta_{13} = 1$ | 3 tightly clustered variables |
| 3 | $\beta_7 = \beta_8 = \beta_{13} = \beta_4 = \beta_5 = \beta_{16} = 1$ | 6 highly correlated variables |
| 4 | $\beta_7 = \beta_8 = \beta_{13} = 1,$ $\beta_4 = \beta_5 = \beta_{16} = 0.1$ | weaker effect for one cluster of variables |
| 5 | $\beta_7 = \beta_8 = \beta_{13} = 0.1,$ $\beta_4 = \beta_5 = \beta_{16} = 1$ | weaker effect for one cluster of variables |
| 6 | $\beta_7 = \beta_8 = \beta_{13} = 1,$ $\beta_4 = \beta_5 = \beta_{16} = -1$ | negative effect for one cluster of variables |
| 7 | $\beta_7 = \beta_8 = \beta_{13} = -1,$ $\beta_4 = \beta_5 = \beta_{16} = 1$ | negative effect for one cluster of variables |
| 8 | $\beta_7 = \beta_8 = \beta_{13} = \beta_2 = \beta_4 = \beta_{14} = 1$ | weakly and strongly correlated variables mixed |
| 9 | $\beta_7 = \beta_8 = \beta_{13} = 1,$ $\beta_2 = \beta_4 = \beta_{14} = 0.1$ | weakly and strongly correlated variables mixed with mixed effects |
| 10 | $\beta_7 = \beta_8 = \beta_{13} = 0.1,$ $\beta_2 = \beta_4 = \beta_{14} = 1$ | weakly and strongly correlated variables mixed with mixed effects |
| 11 | $\beta_7 = \beta_8 = \beta_{13} = 1,$ $\beta_2 = \beta_4 = \beta_{14} = -1$ | weakly and strongly correlated variables mixed with mixed effects |
| 12 | $\beta_7 = \beta_8 = \beta_{13} = -1,$ $\beta_2 = \beta_4 = \beta_{14} = 1$ | weakly and strongly correlated variables mixed with mixed effects |
| 13 | $\beta_7 = \beta_8 = \beta_{13} = \beta_4 = \beta_5 = \beta_{16} = \beta_2 = \beta_{14} = 1$ | mixed scenario |

Supplementary Table S 3: Overview of coefficient structures in real simulation setup. Only non-zero coefficients are specified on a standardized scale.

# 4. Additional results

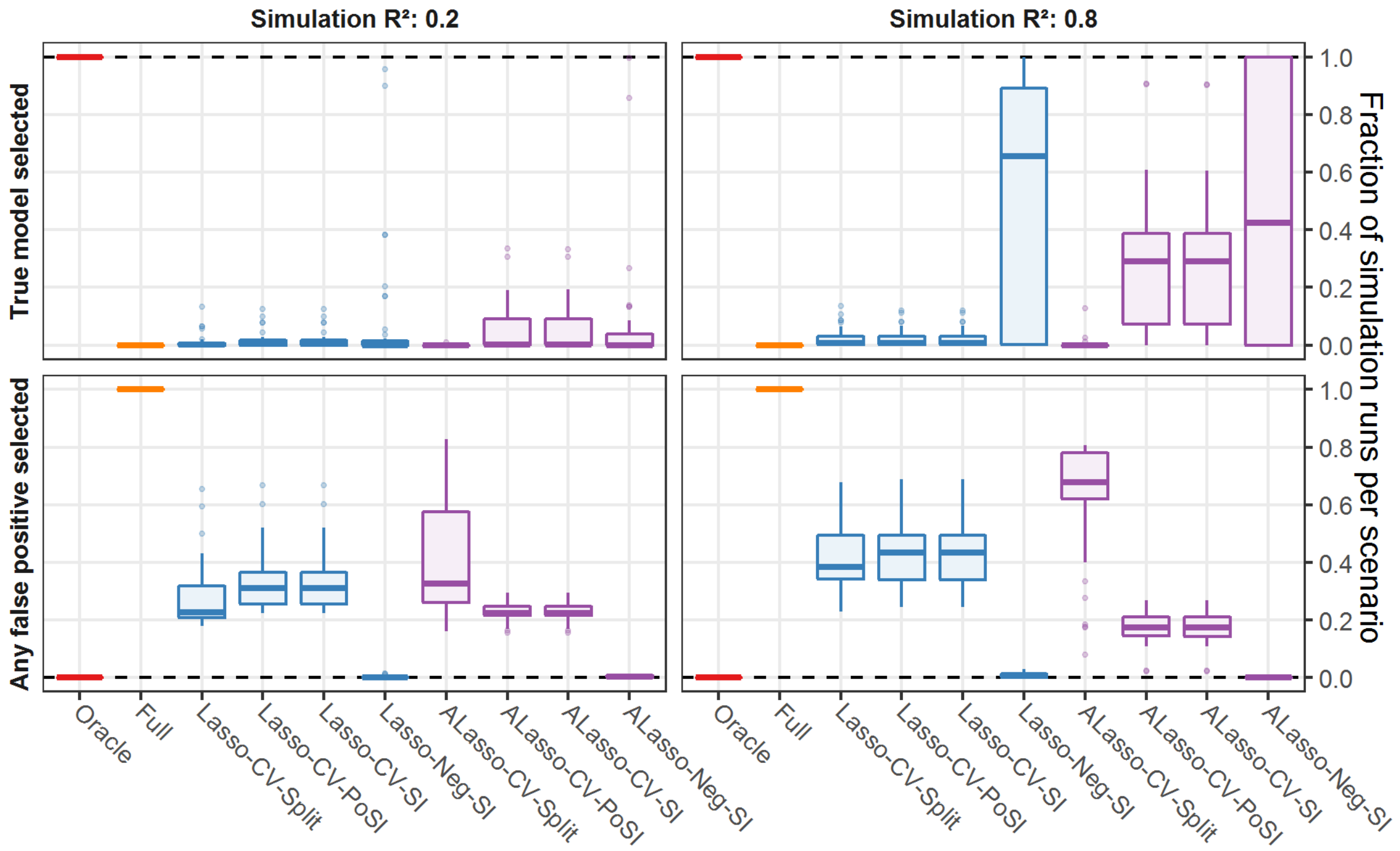

Supplementary Figure S 3: Summary of results from the real setup regarding model selection. For each scenario, we computed the fraction of simulation runs in which the true data generating model (i.e. only true predictors) was selected (top row) and in which any false positive selection (i.e. a true non-predictor was selected) was made (bottom row panel). These results for all scenarios are then summarised by boxplots. The target, optimal values are depicted as dashed lines.

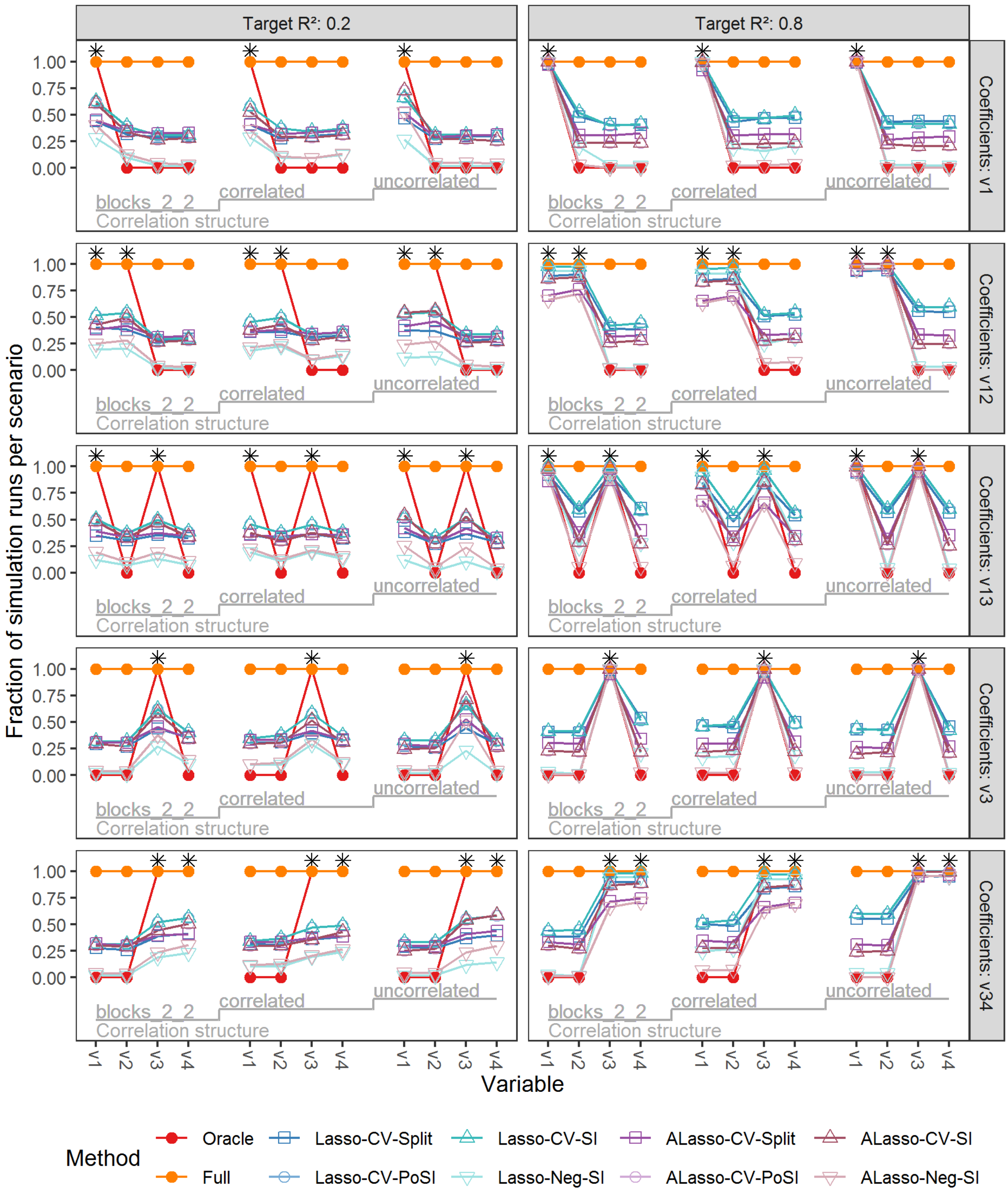

Supplementary Figure S 4: Nested loop plot of individual variable selection frequencies for the toy setup. Only a subset of all scenarios is shown: simulation $R^2$ of 0.5 and sample sizes 40 and 200 are excluded; only a few selected correlation structures (indicated by grey steps; see the Supplementary section 3 for details) and coefficient structures (rows, Supplementary Table S 1) are depicted. Each dot represents the selection frequency for a specific variable over all simulation runs per scenario per method. Black stars indicate which variables have an effect. The red color is used for the Oracle model, which represents perfect selection and represents the full model view target for each variable selection procedure. Similar plots were used to interpret the results from the simulation study. This example plot shows that Lasso-CV methods led to generally higher variable selection frequencies than ALasso-CV across the different simulation scenarios. The Lasso-Neg and ALasso-Neg approaches led to extremely sparse models. Furthermore, in case of correlation (as indicated by the “correlated” or “blocks_2_2” correlation structures) accuracy of selection was generally lower than in the uncorrelated case. Block-correlation also led to slightly different results than overall correlation for certain effect structure (e.g. “v34”).

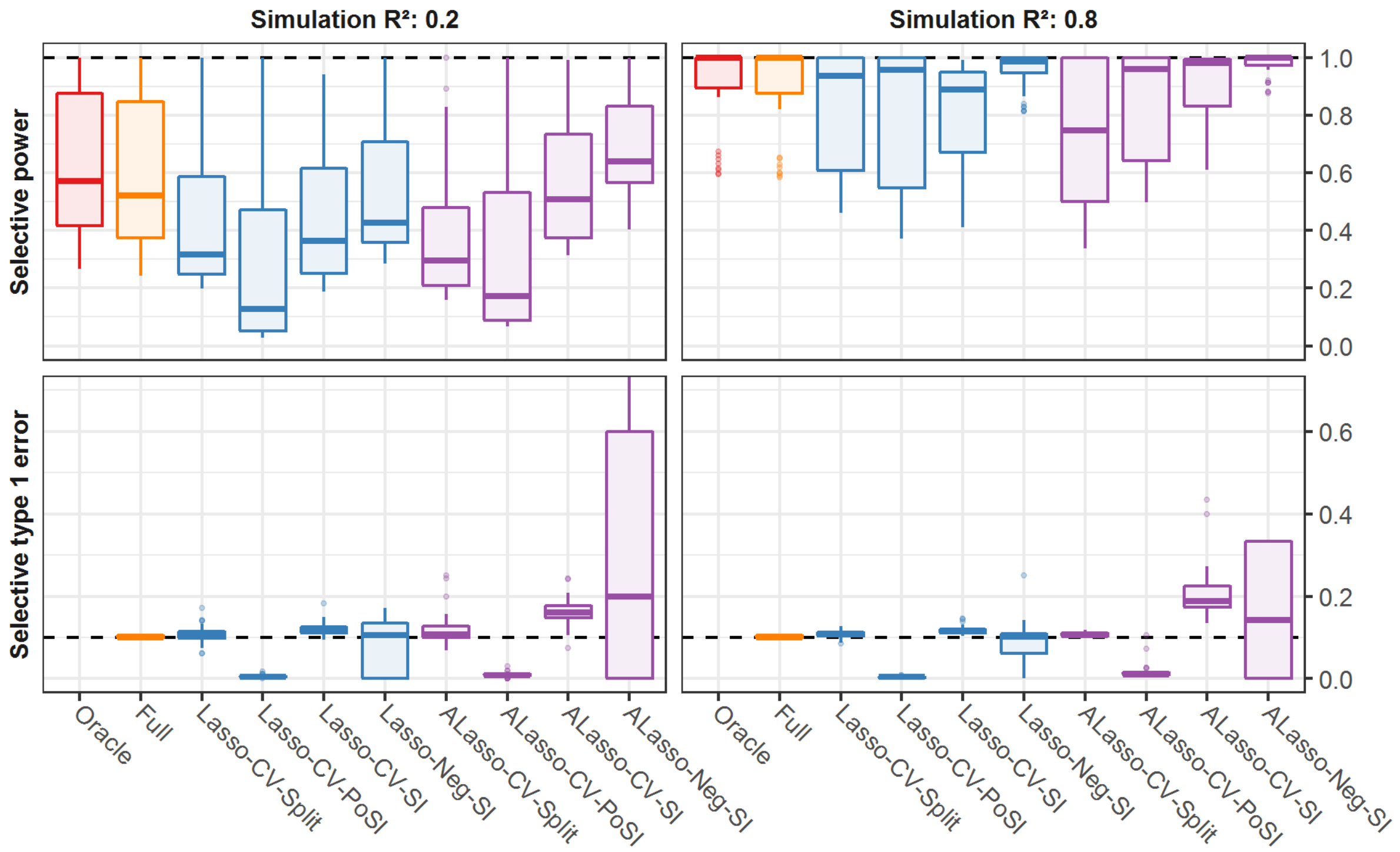

Supplementary Figure S 5: Summary of results from the realistic simulation setup regarding the selective power (top row) and type 1 error (bottom row) of the selective 90% CIs for the submodel inference target (see Table 2). Results for all scenarios are summarised by boxplots. The target values are depicted as dashed lines (1 for power, the nominal significance level of 0.1 for type 1 error). Colors are based on the type of regression model used.

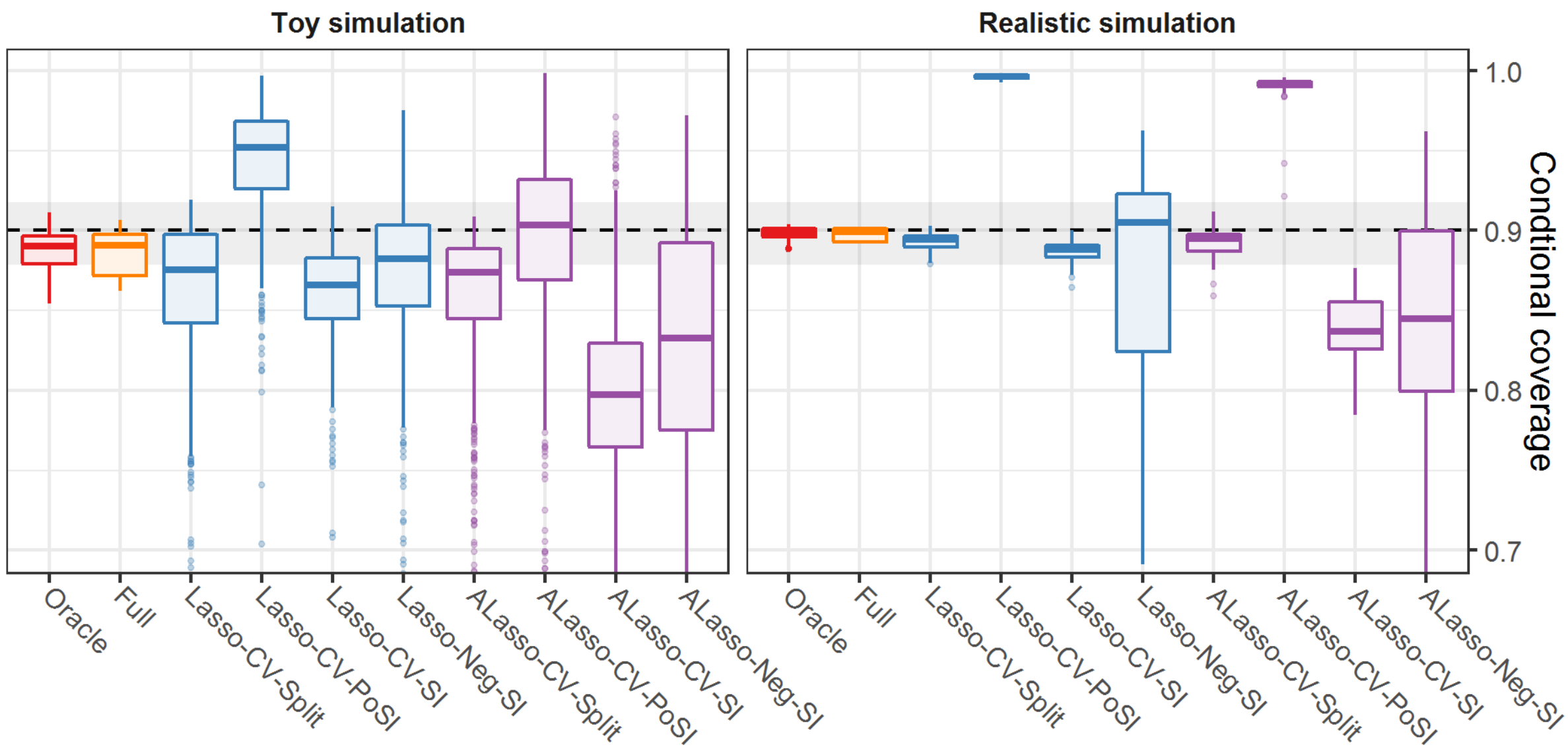

Supplementary Figure S 6: Summary of results from both simulation setups (toy setup in left panel, real setup in right panel) regarding conditional selective coverage of the selective 90% CIs for the submodel inference target (see Table 2). Results are averaged over variables for all scenarios and are summarised by boxplots. The nominal confidence level of 0.9 used in the construction of the CIs is depicted as dashed lines. Colors are based on the type of regression model used. To give an indication of variability expected in this simulation study, the grey areas indicate binomial 95% confidence intervals based on the number of iterations in each scenario (900) and the nominal confidence level.

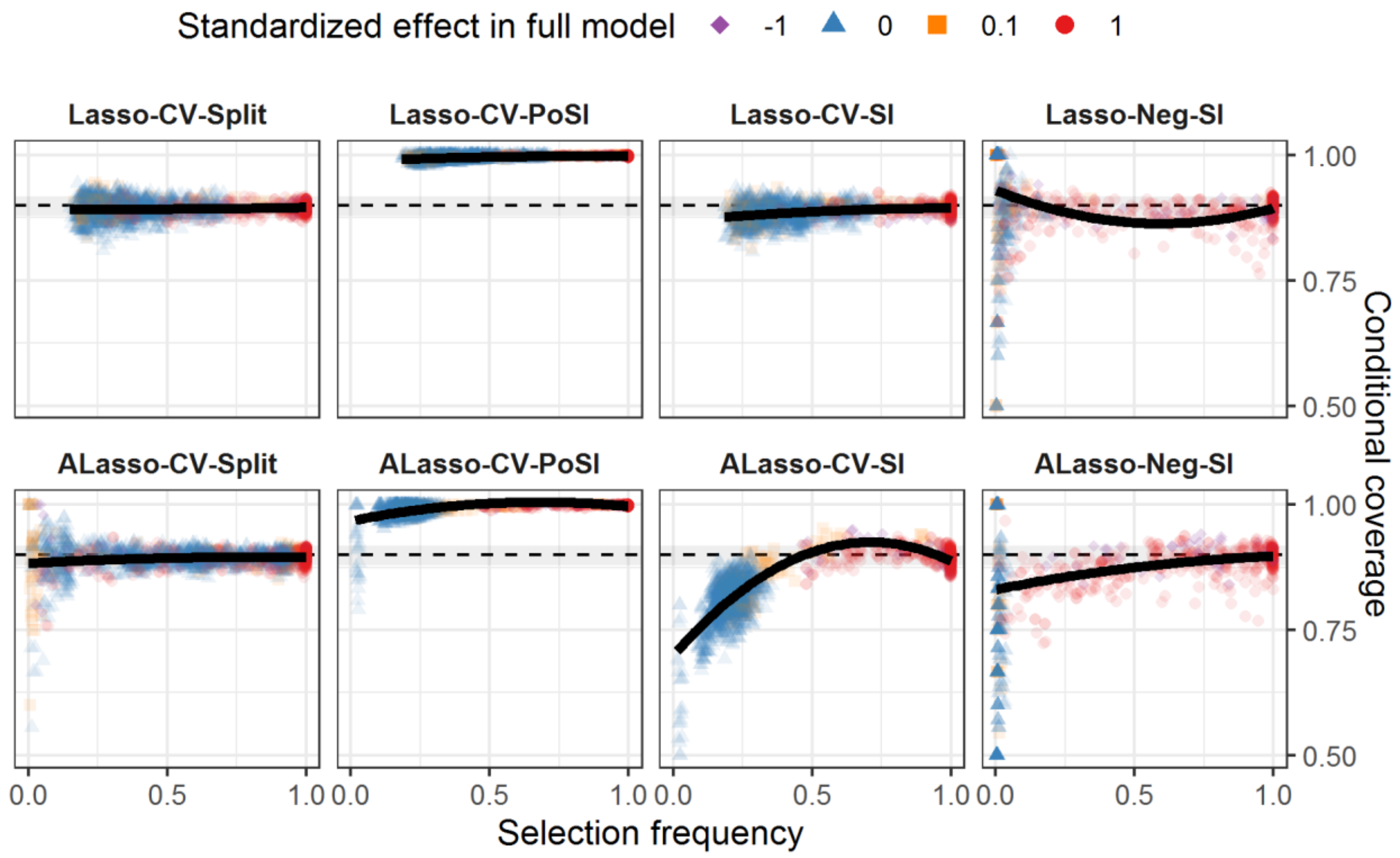

Supplementary Figure S 7: Comparison of selection frequency and conditional coverage from the real simulation setup. Each dot represents results for a single variable in a specific simulation scenario. The target coverage value is depicted as dashed lines. The black line provides a smoothed summary of the observed data (fitted with a quadratic B-spline term with 3 knots for selection frequency). Colors indicate if the variable is a predictor in the full model in the specific scenario. To give an indication of variability expected in this simulation study, the grey areas indicate binomial 95% confidence intervals based on the number of iterations in each scenario (900) and the nominal confidence level.

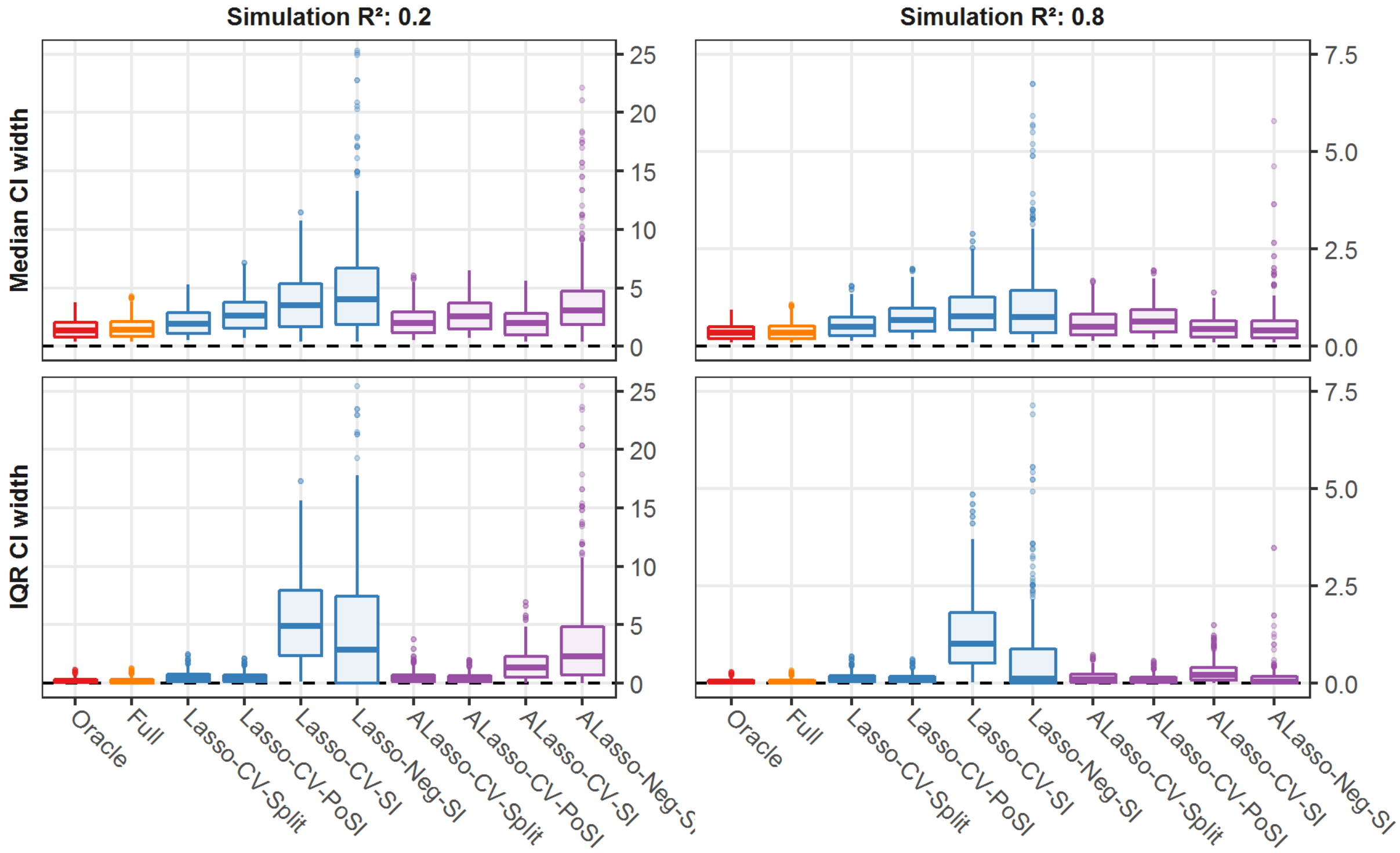

Supplementary Figure S 8: Summary of results from the real simulation setup regarding the width (top row) and variability (bottom row) of the selective 90% CIs for the submodel inference target (see Table 2). Results for all scenarios are averaged over all variables and are summarised by boxplots. Width zero is marked by a dashed lines. Colors are based on the type of regression model used.

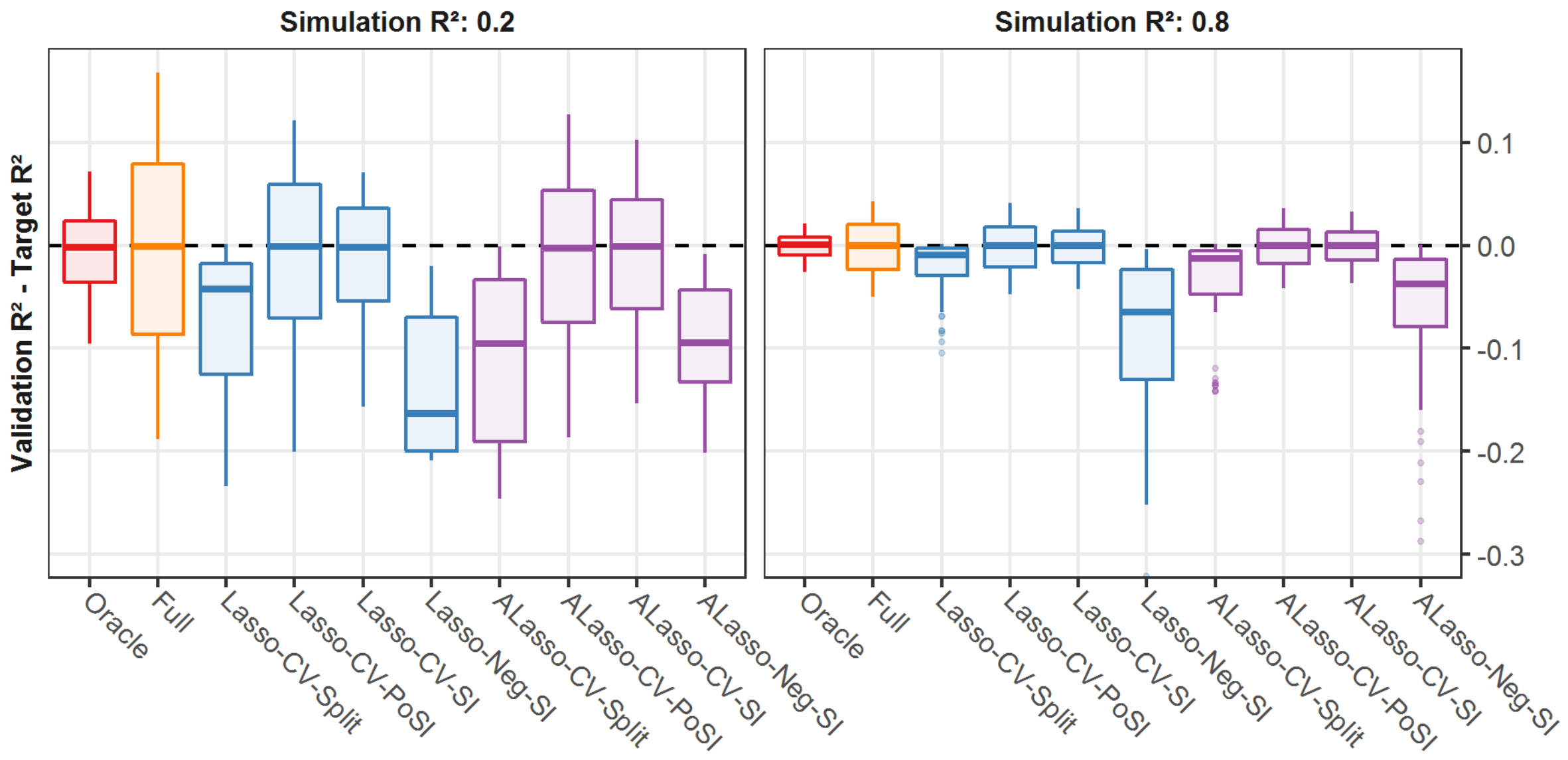

Supplementary Figure S 9: Summary of results from the toy simulation setup regarding predictive accuracy in terms of difference of validation $R^2$ and target $R^2$ (i.e. 0.2 in left panel, 0.8 in right panel). Results for all scenarios are summarised by boxplots. An optimal difference of zero is marked by dashed lines. Colors are based on the type of regression model used.

## 5. R simulation code files

To conform to the publication guidelines on arXiv.org we include the ancillary R example files within this manuscript. Interested readers can copy the code directly into a file to run in the R-software.

### 5.1. Toy_Setup_Demo.R

```
library(Matrix)
library(simdata)

# Helper functions ###################################################
#' @title Correlation matrix with constant off-diagonal elements
#'
#' @param p
#' Number of variables.
#' @param off_diag
#' Values for off diagonal elements.
#'
#' @return
#' Correlation matrix with diagonal elements set to 1.
create_cor_constant <- function(p, off_diag = 0) {
    cmat = matrix(off_diag, nrow = p, ncol = p)
    diag(cmat) = 1
    cmat
}

#' @title Block correlation matrix
#'
#' @param p
#' Number of variables.
#' @param off_diag
#' Values for blocks.
#'
#' @return
#' Correlation matrix with diagonal elements set to 1.
create_cor_block <- function(p, off_diag = rep(0, length(p))) {
    cmat = list()
    for (i in 1:length(p)) {
        cmat[[i]] = matrix(off_diag[i], nrow = p[i], ncol = p[i])
    }
    cmat = bdiag(cmat)
    diag(cmat) = 1
    as.matrix(cmat)
}

# Toy simulationsetup ################################################
# Simple data with 4 variables
# explores many possible structures for 4 variables

# Correlation structures
design_structure = list(
    uncorrelated = mvtnorm_simdesign(
        relations = create_cor_constant(4)
    ),
    correlated = mvtnorm_simdesign(
        relations = create_cor_constant(4, 0.8)
    ),
    correlated_neg = mvtnorm_simdesign(
        relations = create_cor_constant(4, -0.8)
```

```
    ),
    blocks_2_2 = mvtnorm_simdesign(
        relations = create_cor_block(c(2, 2), c(0.8, 0.8))
    ),
    blocks_2_2_neg = mvtnorm_simdesign(
        relations = create_cor_block(c(2, 2), c(0.8, -0.8))
    ),
    blocks_1_3 = mvtnorm_simdesign(
        relations = create_cor_block(c(1, 3), c(0, 0.8))
    ),
    blocks_1_3_neg = mvtnorm_simdesign(
        relations = create_cor_block(c(1, 3), c(0, -0.8))
    )
)

# Coefficient structures
coef_structure = list(
    "v1" = c(1, 0, 0, 0),
    "v12" = c(1, 1, 0, 0),
    "v1234" = c(1, 1, 1, 1),
    "v3" = c(0, 0, 1, 0),
    "v34" = c(0, 0, 1, 1),
    "v13" = c(1, 0, 1, 0),
    "v12_dec" = c(1, 0.1, 0, 0),
    "v34_dec" = c(0, 0, 1, 0.1),
    "v13_dec" = c(1, 0, 0.1, 0),
    "v13_inc" = c(0.1, 0, 1, 0)
)

# set design from which data should be simulated
design = 1
# set coefficients
coef = 1
# set number of observations
n_obs = 100

# simulate data
set.seed(1)
X = simulate_data(design_structure[[design]], n_obs)

# compute linear predictor (no noise, data is assumed to be standardized)
y = X %*% coef_structure[[coef]]
```

## 5.2. Realistic_Setup_Demo.R

```
library(simdata)

# Binder simulationsetup ##############################################
# Uses a slightly modified correlation matrix to achieve stronger
# correlations and more interesting results
# the multiple correlations of the final variables are
#   v1   v2   v3   v4   v5   v6   v7   v8   v9  v10  v11  v12  v13  v14
v15  v16  v17
# 0.63 0.58 0.52 0.44 0.35 0.68 0.66 0.65 0.43 0.46 0.47 0.01 0.71 0.63
0.01 0.61 0.01
# thus one could identify 3 kinds of variables: those with low dependence
# on the rest (v12, 15, 17), those with medium dependence (v3, 4, 5, 9, 10,
11)
# and those with strong dependence (v1, 2, 6, 7, 8, 13, 16)
# in most of these cases, the dependence comes from direct dependence of
# one or two other variables (and not a combination of many)

# Data
relations = cor_from_upper(
    15,
    rbind(c(1,2,0.8), c(1,9,0.5),
          c(3,5,0.5), c(3,9,-0.8),
          c(4,6,-0.8), c(4,7,-0.5),
          c(5,6,-0.5), c(5,12,0.8),
          c(6,7,0.8), c(6,11,0.8), c(6,14,0.5),
          c(7,11,0.5), c(7,14,0.5),
          c(8,9,-0.5), c(8,11,0.5),
          c(11,14,0.8))
)
design_structure = list(
    mvtnorm_simdesign(
        relations = relations,
        transform_initial = function_list(
            v1 = function(z) floor(10 * z[,1] + 55),
            v2 = function(z) z[,2] < 0.6,
            v3 = function(z) exp(0.4 * z[,3] + 3),
            v4 = function(z) z[,4] >= -1.2,
            v5 = function(z) z[,4] >= 0.75,
            v6 = function(z) exp(0.5 * z[,5] + 1.5),
            v7 = function(z) floor(pmax(0, 100 * exp(z[,6]) - 20)),
            v8 = function(z) floor(pmax(0, 80 * exp(z[,7]) - 20)),
            v9 = function(z) z[,8] < -0.35,
            v10 = function(z) (z[,9] >= 0.5) * (z[,9] < 1.5),
            v11 = function(z) z[,9] >= 1.5,
            v12 = function(z) 0.01*floor(100 * (z[,10] + 4)^2),
            v13 = function(z) floor(10 * z[,11] + 55),
            v14 = function(z) floor(10 * z[,12] + 55),
            v15 = function(z) floor(10 * z[,13] + 55),
            v16 = function(z) z[,14] < 0,
            v17 = function(z) z[,15] < 0),
        process_final = list(
            process_truncate = list(
                truncate_multipliers =
                    c(5, NA, 5, NA, NA,
                      5, 5, 5, NA, NA,
                      NA, 5, 5, 5, 5,
                      NA, NA)
            )
        )
```

```
    )
)

# "clusters"
# 1) 3 totally independent vars (12,15,17) - boring
# 2) 3 far apart (2, 4, 14) - quasi independent, but high dependence to
other variables
# 3) 3 highly correlated, positive (7,8,13), otherwise mostly negative
correlation
# 4) 6 clustered variables, positive and negative correlations (3), 4,5,16)
# the remaining variables are "bridges" or intermediates between these
# extremes and less interesting in terms of coefficient structures
# eg. v3, 9, 10, 11 are similar to 3)

# coefficient structures
variable_names = design_structure[[1]]$names_final
# define coefficients such that standardized coefficients are 1
c0 = rep(0, length(variable_names))
names(c0) = variable_names

coef_structure = list()

# 1) cluster 2: 3 quasi independent variables
coef_structure$c2 = c0
coef_structure$c2[c("v2", "v4", "v14")] = 1

# 2) cluster 3: 3 highly positively correlated variables
coef_structure$c3 = c0
coef_structure$c3[c("v7", "v8", "v13")] = 1

# 3) cluster 4: 6 highly correlated variables
coef_structure$c34 = c0
coef_structure$c34[c("v7", "v8", "v13", "v4", "v5", "v16")] = 1
# 4) Weaker effects for one block
coef_structure$c3w4 = coef_structure$c34
coef_structure$c3w4[c("v7", "v8", "v13")] = 0.1
# 5)
coef_structure$c34w = coef_structure$c34
coef_structure$c34w[c("v4", "v5", "v16")] = 0.1
# 6) negative effects
coef_structure$c3neg4 = coef_structure$c34
coef_structure$c3neg4[c("v7", "v8", "v13")] = -1
# 7)
coef_structure$c34neg = coef_structure$c34
coef_structure$c34neg[c("v7", "v8", "v13")] = -1

# 8) cluster 2 + 3 combined
coef_structure$c23 = c0
coef_structure$c23[c("v2", "v4", "v14", "v7", "v8", "v13")] = 1
# 9) weaker effects for one block
coef_structure$c2w3 = coef_structure$c23
coef_structure$c2w3[c("v2", "v4", "v14")] = 0.1
# 10)
coef_structure$c23w = coef_structure$c23
coef_structure$c23w[c("v7", "v8", "v13")] = 0.1
# 11) negative effects
coef_structure$c2neg3 = coef_structure$c23
coef_structure$c2neg3[c("v7", "v8", "v13")] = -1
# 12)
coef_structure$c23neg = coef_structure$c23
coef_structure$c23neg[c("v7", "v8", "v13")] = -1
```

```
# 13) cluster 2 + 4 combined
coef_structure$c234 = c0
coef_structure$c234[c("v2", "v4", "v14", "v7", "v8", "v13", "v5", "v16")] =
1

# set coefficients
coef = 1
# set number of observations
n_obs = 100

# simulate data
set.seed(1)
X = simulate_data(design_structure[[1]], n_obs)

# compute linear predictor (no noise, data is assumed to be standardized)
y = as.matrix(X) %*% coef_structure[[coef]]
```